\documentclass[sigconf,nonacm,screen]{acmart}
\renewcommand\footnotetextcopyrightpermission[1]{}

\usepackage{mathtools}
\usepackage{bm}
\usepackage{booktabs,multirow,array}
\usepackage{graphicx}
\usepackage{subcaption}
\usepackage{algorithm}
\usepackage{algpseudocode}
\usepackage{tikz}
\usetikzlibrary{positioning,fit,arrows.meta,backgrounds,calc,shapes.misc}
\usepackage[capitalise,nameinlink]{cleveref}

\graphicspath{{figures/}}

\newcommand{\method}{GraftQ}
\newcommand{\E}{\mathbb{E}}
\newcommand{\1}{\mathbf{1}}
\newcommand{\R}{\mathbb{R}}
\newcommand{\C}{\mathbb{C}}
\newcommand{\Csrc}{C_{\mathrm{src}}}
\newcommand{\Cr}{C_{\mathrm{ref}}}   
\newcommand{\Cgen}{\hat C}
\newcommand{\Nall}{N_{\mathrm{all}}}
\newcommand{\Ntq}{N_{\mathrm{2q}}}
\newcommand{\NT}{N_{T}}
\newcommand{\clip}{\operatorname{clip}}
\newcommand{\Tr}{\operatorname{Tr}}
\newcommand{\ket}[1]{\lvert #1\rangle}
\newcommand{\bra}[1]{\langle #1\rvert}
\newcommand{\braket}[2]{\langle #1\vert #2\rangle}
\DeclareMathOperator*{\argmin}{arg\,min}

\definecolor{newcomp}{RGB}{226,219,245}
\definecolor{oldcomp}{RGB}{238,238,238}

\begin{document}

\title{Multimodal Generative Quantum Circuit Optimization with Verifier-Guided Reinforcement Learning}

\author{Xiaodi Li}
\authornote{Email: \href{mailto:lixiaodi@fudan.edu.cn}{\nolinkurl{lixiaodi@fudan.edu.cn}}.}
\affiliation{%
  \institution{Intelligent Quantum Inception Inc.}
  \city{Haidian, Beijing}
  \postcode{100085}
  \country{China}}
\affiliation{%
  \institution{iFLYTEK Research}
  \city{Hefei}
  \postcode{230088}
  \country{China}}
\affiliation{%
  \department{Beijing Key Laboratory of Cold Atom Quantum Computation}
  \institution{Tsinghua University}
  \city{Beijing}
  \postcode{100084}
  \country{China}}
\author{Linghui Chen}
\authornote{Corresponding author. Email: \href{mailto:chenlh@mail.ustc.edu.cn}{\nolinkurl{chenlh@mail.ustc.edu.cn}}.}
\affiliation{%
  \institution{iFLYTEK Research}
  \city{Hefei}
  \postcode{230088}
  \country{China}}
\affiliation{%
  \institution{Intelligent Quantum Inception Inc.}
  \city{Haidian, Beijing}
  \postcode{100085}
  \country{China}}
\affiliation{%
  \department{Beijing Key Laboratory of Cold Atom Quantum Computation}
  \institution{Tsinghua University}
  \city{Beijing}
  \postcode{100084}
  \country{China}}
\author{Tao Zhang}
\authornote{Corresponding author. Email: \href{mailto:tao_zhang@mail.tsinghua.edu.cn}{\nolinkurl{tao_zhang@mail.tsinghua.edu.cn}}.}
\affiliation{%
  \department{Institute for Advanced Study}
  \institution{Tsinghua University}
  \city{Beijing}
  \postcode{100084}
  \country{China}}
\affiliation{%
  \department{Beijing Key Laboratory of Cold Atom Quantum Computation}
  \institution{Tsinghua University}
  \city{Beijing}
  \postcode{100084}
  \country{China}}
\affiliation{%
  \institution{Intelligent Quantum Inception Inc.}
  \city{Haidian, Beijing}
  \postcode{100085}
  \country{China}}
\affiliation{%
  \institution{iFLYTEK Research}
  \city{Hefei}
  \postcode{230088}
  \country{China}}
\renewcommand{\shortauthors}{Li et al.}

\begin{abstract}
Quantum circuit optimization replaces a circuit with an equivalent one using fewer costly operations: two-qubit gates, which dominate error on noisy hardware, and non-Clifford $T$ gates, which dominate fault-tolerant overhead. We formulate this task as graph-conditioned sequence-to-sequence generation and introduce \method{} (Graph-Augmented Fusion Transformer for Quantum circuit optimization), a multimodal GNN-Transformer that jointly encodes a source circuit's gate-token sequence and simplified ZX diagram in a Transformer encoder and autoregressively generates an optimized circuit. The model is first pretrained by supervised learning on millions of verified triples of a random source circuit, its simplified ZX diagram, and a target circuit optimized by PyZX. It is then fine-tuned by reinforcement learning: a verifier checks whether each generated circuit is equivalent to the reference circuit, and a reward function scores its gate-count reduction relative to that reference. We compare Group Relative Policy Optimization (GRPO) with Maximum-Likelihood Reinforcement Learning (MaxRL), whose binary-reward objective we generalize to continuous non-negative rewards, and propose a Hybrid advantage estimator that combines the strengths of both. A size curriculum further extends training from short (20--100-gate) circuits up to 300 gates on 5--20 qubits. On held-out benchmarks, verifier-guided RL with curriculum training raises equivalence accuracy (pass@8) on 20--100-gate circuits from $67.7\%$ under supervised learning to $94.1\%$ while surpassing PyZX in two-qubit gate reduction and nearly matching its $T$-gate counts, and scales optimization to 300-gate circuits while generalizing to deeper unseen circuits.
\end{abstract}


\maketitle

\section{Introduction}
\label{sec:intro}

Quantum computers promise speedups for problems such as integer factoring~\cite{shor1997polynomial}, unstructured search~\cite{grover1996fast}, and the simulation of quantum many-body systems and molecules~\cite{nielsen2010quantum}. Over the past decade the hardware has moved from proof-of-principle devices to processors with tens to hundreds of qubits. The first logical-qubit experiments have demonstrated surface-code memories below the error-correction threshold~\cite{google2025below} and programmable logical processors on neutral atoms~\cite{bluvstein2024logical}. At the same time, resource estimates for cryptographically relevant algorithms keep falling as algorithms and compilation improve~\cite{gidney2021factor,gidney2025factor}. In both the noisy intermediate-scale quantum (NISQ) regime~\cite{preskill2018nisq} and the emerging fault-tolerant regime, the gate count and gate composition of a compiled circuit largely determine whether execution is feasible in practice.

Turning an algorithm into operations on a device is the task of quantum compilation. A compiler translates a hardware-independent circuit into an executable one: it decomposes high-level operations into a native gate set, maps logical to physical qubits and inserts routing operations to respect connectivity~\cite{li2019tackling,zulehner2019efficient}, and applies semantics-preserving optimizations throughout. Production toolchains such as Qiskit~\cite{javadiabhari2024qiskit}, $\mathrm{t}|\mathrm{ket}\rangle$~\cite{sivarajah2021tket}, and PyZX~\cite{kissinger2020pyzx}, together with formally verified optimizers such as VOQC~\cite{hietala2021voqc}, implement these stages as pipelines of passes. Their cost models depend on the regime. On NISQ hardware, two-qubit gates incur substantially higher error rates than single-qubit gates, so the two-qubit gate count largely determines circuit fidelity. Under fault tolerance, non-Clifford $T$ gates are intrinsically more resource-intensive than Clifford operations: the Eastin--Knill theorem precludes transversal $T$ in most codes~\cite{eastin2009restrictions}, so $T$ gates generally require code switching~\cite{anderson2014fault} or magic state preparation techniques~\cite{bravyi2005universal,litinski2019magic,gidney2024cultivation}, making $T$-count a dominant cost metric.

Within this compilation stack, \emph{circuit optimization} seeks a circuit that implements the same unitary as the input, up to a global phase, while using fewer costly gates. This problem is computationally intractable: exact optimization is NP-hard in general, and even approximate optimization remains hard under natural cost metrics~\cite{vandewetering2024optimising}. Moreover, circuit representations are highly non-unique: commuting gates can be reordered, identities can be inserted or cancelled anywhere, and a single unitary admits exponentially many implementations, so simplifications that are invisible locally may only emerge once distant gates are brought together. Every reduction in gate count directly lowers execution error on noisy hardware or resource overhead under fault tolerance, making circuit optimization a critical stage of quantum compilation.

Existing approaches to circuit optimization include rule-based rewriters and heuristic search methods~\cite{amy2014polynomial,nam2018automated,sivarajah2021tket,hietala2021voqc,xu2022quartz,xu2023queso,pointing2024quanto,xu2025guoq}, ZX-calculus simplification~\cite{coecke2011interacting,duncan2020graph,kissinger2020reducing,vandewetering2020zx}, reinforcement-learning methods that select sequential transformations~\cite{fosel2021quantum,li2024quarl,kremer2024practical,nagele2024optimizing,riu2025reinforcement,mattick2025optimizing,ruiz2025alphatensor}, and generative models that synthesize an optimized circuit directly, among others. Generative models offer a key structural advantage: by synthesizing an optimized circuit end-to-end, they can directly capture global circuit structure. Transformers, in particular, process the entire input globally so that each output gate can depend on arbitrarily distant gates in the source. Diffusion and flow-based models synthesize circuits from target unitaries~\cite{furrutter2024diffusion} or transport source circuits to optimized equivalents~\cite{hofstetter2026bridge}, while large language models have been fine-tuned to generate circuits~\cite{jern2025agentq}. However, purely generative approaches also face significant challenges: for autoregressive models, the accuracy of generating equivalent circuits drops rapidly as circuit length grows~\cite{saeedi2026close}, and without an explicit correctness signal their output validity is difficult to guarantee.

In this work, we adopt an encoder--decoder Transformer framework for quantum circuit optimization. A circuit is naturally represented as a sequence of gate tokens (e.g., \texttt{HAD\_2}, \texttt{T\_5}, or \texttt{CNOT\_3\_7}) drawn from a finite vocabulary, similar to gate statements in OpenQASM~\cite{cross2017openqasm}. Source and target circuits share this vocabulary, casting circuit optimization as a sequence-to-sequence (Seq2Seq) task. 
However, a plain Seq2Seq model is insufficient. First, a flat token sequence obscures the two-dimensional qubit connectivity of the source circuit. Second, autoregressive decoding suffers from error accumulation under supervised training alone~\cite{saeedi2026close}.

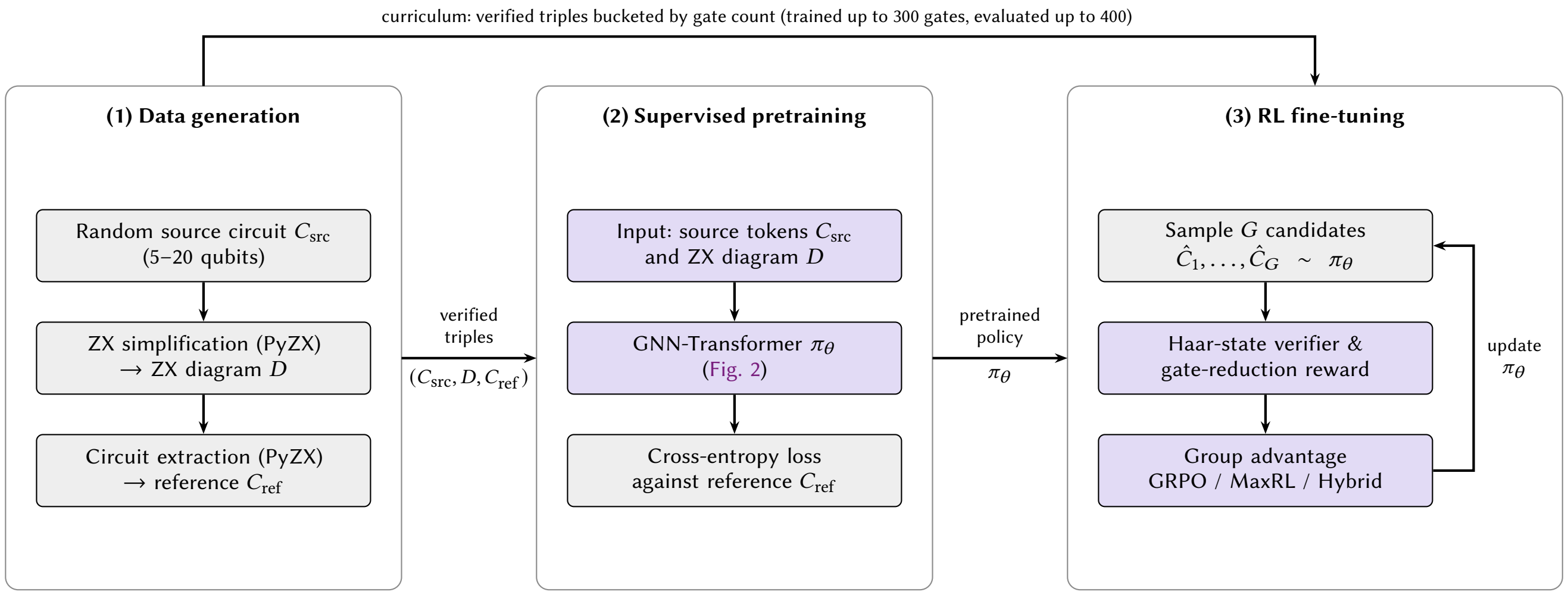
\begin{figure*}[t]
\centering
\begin{tikzpicture}[
  font=\sffamily\footnotesize,
  >={Stealth[length=4pt,width=3pt]},
  box/.style={draw, rounded corners=2pt, minimum height=0.8cm,
              text width=3.5cm, align=center, inner sep=3pt},
  new/.style={box, fill=newcomp},
  old/.style={box, fill=oldcomp},
  arr/.style={->, thick},
  panel/.style={draw=black!45, rounded corners=4pt},
  ttl/.style={font=\sffamily\footnotesize\bfseries},
  lbl/.style={font=\sffamily\scriptsize, align=center},
]

\draw[panel] (-2.2,1.15) rectangle (2.2,-4.45);
\node[ttl] at (0,0.8) {(1) Data generation};
\node[old] (a1) at (0,-0.625) {Random source circuit $\Csrc$\\(5--20 qubits)};
\node[old] (a2) at (0,-1.875) {ZX simplification (PyZX)\\$\to$ ZX diagram $D$};
\node[old] (a3) at (0,-3.125) {Circuit extraction (PyZX)\\$\to$ reference $\Cr$};
\draw[arr] (a1) -- (a2);
\draw[arr] (a2) -- (a3);

\draw[panel] (3.7,1.15) rectangle (8.1,-4.45);
\node[ttl] at (5.9,0.8) {(2) Supervised pretraining};
\node[new] (b1) at (5.9,-0.625) {Input: source tokens $\Csrc$\\and ZX diagram $D$};
\node[new] (b2) at (5.9,-1.875) {GNN-Transformer $\pi_\theta$\\(\cref{fig:model})};
\node[old] (b3) at (5.9,-3.125) {Cross-entropy loss\\against reference $\Cr$};
\draw[arr] (b1) -- (b2);
\draw[arr] (b2) -- (b3);

\draw[panel] (9.6,1.15) rectangle (15.1,-4.45);
\node[ttl] at (12.35,0.8) {(3) RL fine-tuning};
\node[old] (c1) at (11.8,-0.625) {Sample $G$ candidates\\$\hat C_1,\dots,\hat C_G\sim\pi_\theta$};
\node[new] (c2) at (11.8,-1.875) {Haar-state verifier \&\\gate-reduction reward};
\node[new] (c3) at (11.8,-3.125) {Group advantage\\GRPO / MaxRL / Hybrid};
\draw[arr] (c1) -- (c2);
\draw[arr] (c2) -- (c3);
\draw[arr] (c3.east) -- ++(0.45,0) |- (c1.east);
\node[lbl, anchor=west] at (14.15,-1.875) {update\\$\pi_\theta$};

\draw[arr] (2.2,-1.875) -- (3.7,-1.875)
  node[lbl, midway, above] {verified\\triples}
  node[lbl, midway, below] {$(\Csrc,D,\Cr)$};
\draw[arr] (8.1,-1.875) -- (9.6,-1.875)
  node[lbl, midway, above] {pretrained\\policy}
  node[lbl, midway, below] {$\pi_\theta$};
\draw[arr] (0,1.15) -- (0,1.7) -| (12.35,1.15);
\node[lbl, above] at (6.15,1.7)
  {curriculum: verified triples bucketed by gate count (trained up to 300 gates, evaluated up to 400)};

\end{tikzpicture}
\caption{\method{} pipeline. Components introduced in this work are shaded lavender; standard steps and existing tools are shaded gray. \textbf{(1)} A random source circuit is simplified into a ZX diagram and extracted by PyZX to produce a reference circuit $\Cr$. \textbf{(2)} A GNN-Transformer takes both the source token sequence $\Csrc$ and the simplified ZX diagram $D$ as input and is pretrained to generate $\Cr$. \textbf{(3)} The pretrained policy is fine-tuned by RL: it samples $G$ candidate circuits per input, a random-state verifier tests whether each candidate is equivalent to $\Cr$, a reward function scores gate reduction relative to $\Cr$, and GRPO, continuous MaxRL, or the Hybrid advantage estimator updates the policy across a curriculum of increasing circuit sizes.}
\label{fig:pipeline}
\end{figure*}

We address both limitations with \method{} and a post-training framework (\cref{fig:pipeline}; headline results in \cref{tab:headline}) built on the following contributions:
\begin{itemize}
  \item \textbf{A multimodal GNN-Transformer} (\S\ref{sec:model}). Each source circuit is paired with its simplified ZX diagram, which captures the topological structure of the original circuit. A graph convolutional network embeds the nodes of the diagram, and its node embeddings are concatenated with the gate-token embeddings of the original circuit. A shared Transformer encoder then processes both modalities before an autoregressive decoder generates the optimized circuit.

  \item \textbf{RL with an automatic verifier} (\S\ref{sec:verify}--\ref{sec:rl}). Circuit equivalence can be verified automatically, placing circuit optimization alongside mathematical reasoning and code generation, where verifiable rewards have enabled substantial gains through reinforcement learning~\cite{le2022coderl,shao2024deepseekmath,guo2025deepseekr1}. Hence, we follow the pretrain-then-RL approach of modern language models~\cite{ouyang2022training} and fine-tune the pretrained model with a random-state verifier, which propagates a single Haar-random input state through each generated circuit and the reference circuit to test whether they are equivalent. The reward combines output-state fidelity (for incorrect candidates) with gate-count reduction relative to the reference (for verified candidates), strictly prioritizing equivalence over gate reduction.

  \item \textbf{Continuous MaxRL and a Hybrid advantage estimator} (\S\ref{sec:rl}, Appendices~\ref{app:maxrl}--\ref{app:hybrid}). We adapt GRPO~\cite{shao2024deepseekmath} and MaxRL~\cite{tajwar2026maxrl} to this graded reward, extending MaxRL from binary to continuous non-negative rewards. We further propose a Hybrid estimator that applies MaxRL weighting to groups containing incorrect candidates and GRPO normalization to all-correct groups, automatically shifting each input from learning equivalence to reducing gate count as accuracy improves.

  \item \textbf{Curriculum learning over circuit size} (\S\ref{sec:data}--\ref{sec:setup}). To further alleviate autoregressive drift, we train on millions of verified (source circuit, ZX diagram, PyZX reference) triples on 5--20 qubits with a two-stage curriculum over circuit size. Starting from circuits with 20--100 gates, the curriculum progressively extends training through 101--200 to 201--300 gates without degrading short-circuit accuracy, and we evaluate length generalization on unseen circuits with 301--400 gates.
\end{itemize}

\begin{table}[t]
\centering
\small
\setlength{\tabcolsep}{4.2pt}
\caption{Headline comparison on 9{,}000 held-out 20--100-gate circuits (5--20 qubits, $K=8$, $\mathcal T=0.2$). Gate reductions are relative to the source on the three-model common success set ($n_{\mathrm{com}}=5{,}574$; full breakdown in \cref{tab:main}); \emph{better (2Q)} is the fraction of each model's solved circuits with strictly fewer 2Q gates than PyZX (\S\ref{sec:res-main}).}
\label{tab:headline}
\begin{tabular}{@{}lccccc@{}}
\toprule
 & & \multicolumn{3}{c}{gates reduced} & \\
\cmidrule(lr){3-5}
System & pass@$8$ & all & 2Q & $T$ & better (2Q) \\
\midrule
PyZX reference~\cite{kissinger2020pyzx} & -- & 14.35 & 0.84 & \textbf{6.26} & -- \\
\method{} (SFT) & 67.7\% & 14.39 & 0.84 & \textbf{6.26} & 0.1\% \\
\method{} (SFT + Hybrid RL) & 83.5\% & 15.71 & \textbf{1.07} & 6.25 & 21.8\% \\
\method{} (+ Curriculum) & \textbf{94.1\%} & \textbf{15.81} & \textbf{1.07} & 6.24 & \textbf{23.6\%} \\
\bottomrule
\end{tabular}
\end{table}

\section{Background and Related Work}
\label{sec:background}

\subsection{Quantum circuits and their optimization}
\label{sec:bg-circuits}

An $n$-qubit pure state is a unit vector $\ket{\psi}\in\C^{d}$ with $d=2^n$, and a quantum gate acts on it by a unitary matrix~\cite{nielsen2010quantum}. A circuit $C=(g_1,\dots,g_m)$ is a sequence of gates, each acting on one or more qubits, whose combined action is the $d\times d$ unitary $U(C)=U(g_m)\cdots U(g_1)$. We work with the gate set
\begin{equation}
\mathcal{G}=\{H,\;X,\;Z,\;S,\;T,\;R_Z(\ell\pi/4),\;\mathrm{CNOT},\;\mathrm{CZ}\},
\label{eq:gateset}
\end{equation}
where $R_Z(\alpha)=\mathrm{diag}(1,e^{i\alpha})$ is a $Z$-axis phase rotation and $\ell\in\{1,2,3,5,6,7\}$. This gate set covers both our source circuits and the circuits extracted by PyZX~\cite{kissinger2020pyzx}. Up to global phase, $S=R_Z(\pi/2)$, $T=R_Z(\pi/4)$, and $Z=R_Z(\pi)$. Hadamard, $S$, and CNOT generate the Clifford group, and adding the non-Clifford $T$ gate makes the set approximately universal (Clifford$+T$)~\cite{bravyi2005universal}. We count as non-Clifford every $T$ gate and every phase whose angle is an odd multiple of $\pi/4$.

Two circuits are \emph{equivalent}, written $C\equiv C'$, when their unitaries differ by at most a global phase:
\begin{equation}
C\equiv C' \iff \exists\,\varphi:\; U(C)=e^{i\varphi}\,U(C').
\label{eq:equiv}
\end{equation} Let $\Nall(C)$, $\Ntq(C)$, and $\NT(C)$ denote the total, two-qubit, and non-Clifford gate counts, and let $\kappa(C)=w_{\mathrm{all}}\Nall(C)+w_{\mathrm{2q}}\Ntq(C)+w_T\NT(C)$ be a cost with non-negative weights. \method{} targets
\begin{equation}
C^\star\in\argmin_{C'\in\mathcal{G}^\ast:\;C'\equiv C}\;\kappa(C'),
\label{eq:opt}
\end{equation}
where $\mathcal{G}^\ast$ is the set of finite circuits over $\mathcal{G}$ on the same $n$ qubits. Optimizing individual gate counts in Clifford$+T$ circuits is NP-hard~\cite{vandewetering2024optimising}. Even checking a candidate is hard: deciding worst-case approximate equivalence in operator norm (up to global phase) is QMA-complete~\cite{janzing2005non}, and a dense comparison of unitaries needs $O(4^n)$ memory. We therefore study a learned, amortized approach: a conditional generator proposes candidates, a verifier checks equivalence, and reinforcement learning rewards lower $\kappa$. Because the optimum of \cref{eq:opt} is unknown, we measure progress against a PyZX reference $\Cr\equiv C$ using
\begin{equation}
\Delta_k(C')=\frac{N_k(\Cr)-N_k(C')}{N_k(\Cr)},\qquad k\in\{\mathrm{all},\mathrm{2q},T\},
\label{eq:delta}
\end{equation}
where positive values mean the generated circuit uses fewer gates than the reference.  

The ZX-calculus represents a circuit as a graph of \emph{spiders} with phases, connected by plain or Hadamard edges~\cite{coecke2011interacting,vandewetering2020zx}. Graphical rewrite rules preserve semantics and can remove large numbers of spiders, revealing global structure that a flat gate sequence obscures. Extracting a simplified diagram back into a circuit, however, is efficient when the diagram has generalized flow~\cite{backens2021there} but \#P-hard in general~\cite{debeaudrap2022circuit}, and extraction can introduce additional CNOT gates~\cite{staudacher2023reducing}. Rather than relying solely on direct circuit extraction as the final output, \method{} conditions on the simplified diagram alongside the source circuit.

\subsection{Related work on circuit optimization}
\label{sec:bg-related}

Rule-based optimizers apply gate cancellation, commutation, peephole rewrites, and template matching~\cite{amy2014polynomial,nam2018automated,sivarajah2021tket,hietala2021voqc}, while search-based superoptimizers such as Quartz, QUESO, Quanto, and GUOQ combine synthesized rewrite rules with heuristic search to explore equivalent circuits~\cite{xu2022quartz,xu2023queso,pointing2024quanto,xu2025guoq}. ZX-calculus methods~\cite{duncan2020graph,kissinger2020reducing} are among the strongest $T$-count optimizers.

Reinforcement-learning methods learn which circuit transformation to apply next. Fösel et al.\ apply deep RL to local circuit rewrites~\cite{fosel2021quantum}. Quarl learns a GNN policy over Quartz rewrites~\cite{li2024quarl}. Several works act on ZX diagrams: GNN policies trained with PPO~\cite{schulman2017ppo} choose ZX rewrites~\cite{riu2025reinforcement,nagele2024optimizing}, and tree search with GNNs finds CNOT-reducing combinations~\cite{mattick2025optimizing}. Kremer et al.\ train RL agents for Clifford-circuit synthesis and transpilation~\cite{kremer2024practical}, and AlphaTensor-Quantum reduces $T$-count through tensor decompositions~\cite{ruiz2025alphatensor}. These methods preserve equivalence by construction, but are bounded by their action libraries and rewrite-trajectory length.

A third line of work generates optimized circuits directly. Charton et al.\ train Transformers to rewrite ZX diagrams~\cite{charton2023teaching}. Diffusion models synthesize circuits from a target unitary~\cite{furrutter2024diffusion} or transport source circuits to optimized equivalents~\cite{hofstetter2026bridge}. LLMs have been fine-tuned to write and optimize quantum programs~\cite{dupuis2024qiskit,jern2025agentq}. Closest to our setting, Saeedi et al.\ find that the equivalence accuracy of autoregressive circuit generation drops sharply as output length grows~\cite{saeedi2026close}.

Specifically, autoregressive sequence-to-sequence (Seq2Seq) models map an input sequence $x$ to an output sequence $y=(y_1,\dots,y_L)$ via
\begin{equation}
\pi_\theta(y\mid x)=\prod_{t=1}^{L}\pi_\theta(y_t\mid y_{<t},x),
\label{eq:ar}
\end{equation}
parameterized here by an encoder--decoder Transformer~\cite{vaswani2017attention}. Tokenizing gates on explicit qubits casts circuit optimization into this framework, where each input circuit admits many equivalent target sequences. Standard maximum-likelihood training against a single reference circuit, however, optimizes token-level cross-entropy rather than unitary equivalence or gate cost, penalizes alternative lower-cost implementations, and relies solely on a one-dimensional input sequence. These limitations motivate conditioning the encoder on ZX graph structure and post-training the generator with verifier-guided reinforcement learning (\S\ref{sec:method}).

\section{Model and Training}
\label{sec:method}

\method{} is a conditional generator $\pi_\theta(y\mid x)$. The input $x=(\Csrc,D)$ pairs the token sequence of a source circuit with a ZX diagram $D$ of that circuit. The output $y$ is the token sequence of a candidate optimized circuit $\Cgen$. This section describes how circuits and diagrams are encoded (\S\ref{sec:repr}), the GNN-Transformer (\S\ref{sec:model}), masked decoding (\S\ref{sec:decode}), the verifier and reward (\S\ref{sec:verify}), supervised pretraining (\S\ref{sec:sft}), and reinforcement learning with GRPO, MaxRL, and Hybrid advantages (\S\ref{sec:rl}).

\subsection{Circuit and diagram representation}
\label{sec:repr}

For the circuit modality, each token represents one gate acting on explicit qubit indices: single-qubit gates \texttt{NAME\_q}, two-qubit gates \texttt{NAME\_c\_t} (control, then target), and phase rotations \texttt{RZ(l/4)\_q}. Combining $\{H,X,S,T,Z\}$ (with $H,X$ written \texttt{HAD}, \texttt{NOT}) and $R_Z(\ell\pi/4)$ ($\ell\in\{1,2,3,5,6,7\}$, reduced to lowest terms) across up to $n_{\max}=20$ qubits, CNOT and CZ on ordered qubit pairs, and four framing symbols (\texttt{<pad>}, \texttt{<unk>}, \texttt{<bos>}, \texttt{<eos>}) yields $|\mathcal{V}|=984$ tokens (retaining overlapping gate notations so both $\Csrc$ and PyZX outputs map directly without rewrites). Source and target circuits share this vocabulary and are wrapped as \texttt{<bos>}\,$g_1\cdots g_m$\,\texttt{<eos>}; for instance, the three-qubit fragment $H_0\,\mathrm{CNOT}_{0,1}\,\allowbreak T_1\,\allowbreak R_Z(3\pi/4)_2\,\allowbreak\mathrm{CZ}_{1,2}$ is tokenized as
\begin{equation*}
  \text{\small\texttt{<bos> HAD\_0 CNOT\_0\_1 T\_1 RZ(3/4)\_2 CZ\_1\_2 <eos>}}.
\end{equation*}
Unlike subword-tokenized QASM text, this gate-level vocabulary keeps each gate and its qubits in a single symbol, sets sequence length to gate count plus two, and supports exact qubit-range masks (\S\ref{sec:decode}).

Each vertex $v$ of the simplified ZX diagram $D=(V,E)$ (\S\ref{sec:data}) is represented by a $(2n_{\max}+5)$-dimensional feature vector
\begin{equation}
\bm{f}_v=\big[\underbrace{\bm{e}_{\mathrm{type}(v)}}_{2}\;\big\|\;\underbrace{\bm{b}_v}_{2n_{\max}+1}\;\big\|\;\underbrace{\sin\alpha_v,\ \cos\alpha_v}_{2}\big],
\label{eq:node}
\end{equation}
concatenating a spider-type one-hot vector $\bm{e}_{\mathrm{type}(v)}$ (distinguishing Z spiders from other vertices), a boundary one-hot vector $\bm{b}_v$ (identifying input qubit $q$, output qubit $q$, or an internal spider), and the sine and cosine of the spider phase $\alpha_v$.

\subsection{GNN-Transformer}
\label{sec:model}

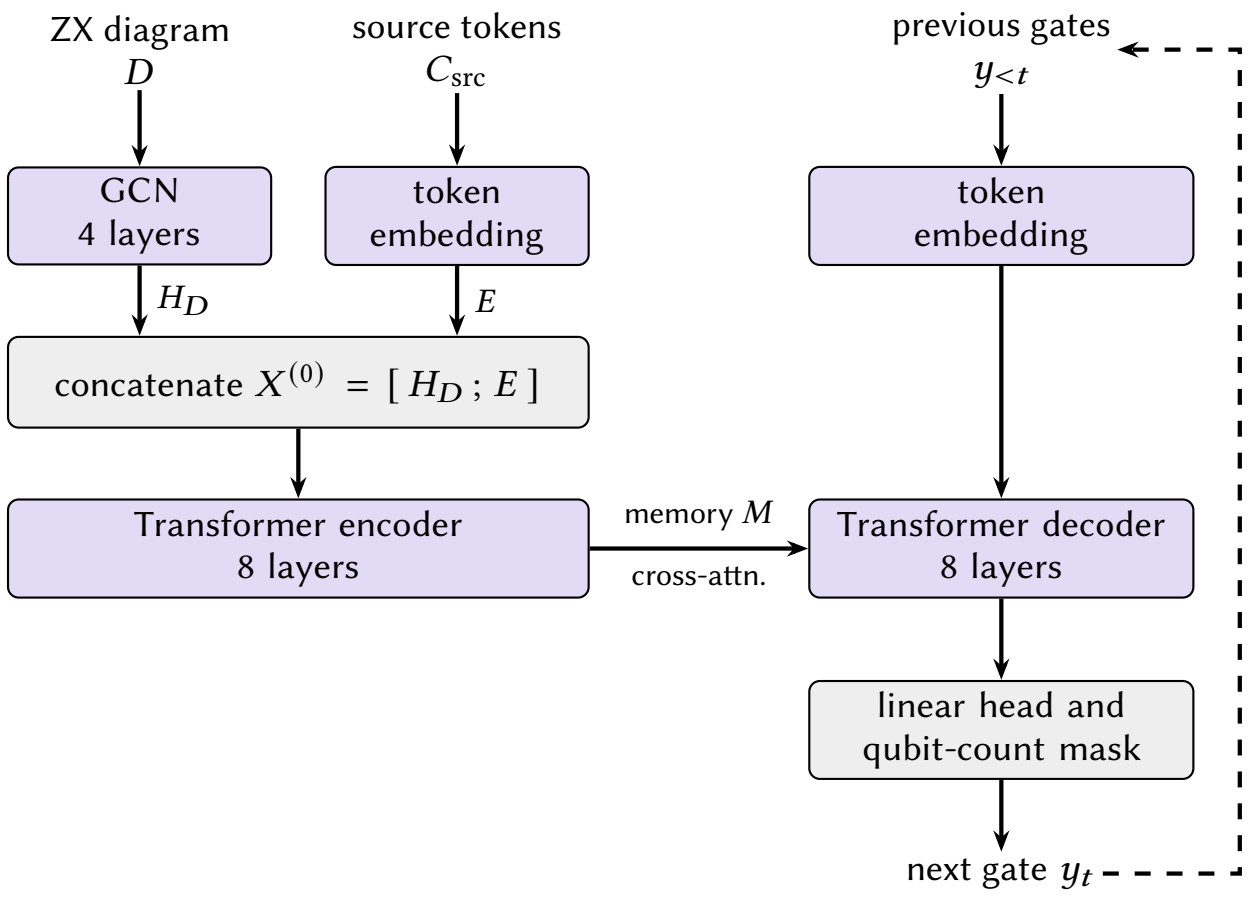
\begin{figure}[t]
\centering
\begin{tikzpicture}[
  font=\sffamily\footnotesize,
  >={Stealth[length=4pt,width=3pt]},
  box/.style={draw, rounded corners=2pt, minimum height=0.6cm,
              align=center, inner sep=2pt},
  mod/.style={box, fill=newcomp},
  op/.style={box, fill=oldcomp},
  io/.style={inner sep=1pt, align=center},
  arr/.style={->, thick},
  lbl/.style={font=\sffamily\scriptsize, align=center},
]

\node[io] (inD) at (0.9,0) {ZX diagram\\$D$};
\node[io] (inS) at (3.0,0) {source tokens\\$\Csrc$};
\node[mod, text width=1.6cm] (gcn) at (0.9,-1.1) {GCN\\4 layers};
\node[mod, text width=1.6cm] (emb) at (3.0,-1.1) {token\\embedding};
\node[op,  text width=3.7cm] (cat) at (1.95,-2.2) {concatenate $X^{(0)}=[\,H_D\,;\,E\,]$};
\node[mod, text width=3.7cm] (enc) at (1.95,-3.3) {Transformer encoder\\8 layers};
\draw[arr] (inD) -- (gcn);
\draw[arr] (inS) -- (emb);
\draw[arr] (gcn) -- node[lbl, right] {$H_D$} (gcn |- cat.north);
\draw[arr] (emb) -- node[lbl, right] {$E$} (emb |- cat.north);
\draw[arr] (cat) -- (enc);

\node[io] (inY) at (6.6,0) {previous gates\\$y_{<t}$};
\node[mod, text width=2.4cm] (demb) at (6.6,-1.1) {token\\embedding};
\node[mod, text width=2.4cm] (dec)  at (6.6,-3.3) {Transformer decoder\\8 layers};
\node[op,  text width=2.4cm] (head) at (6.6,-4.5) {linear head and\\qubit-count mask};
\node[io] (out) at (6.6,-5.45) {next gate $y_t$};
\draw[arr] (inY) -- (demb);
\draw[arr] (demb) -- (dec);
\draw[arr] (dec) -- (head);
\draw[arr] (head) -- (out);

\draw[arr] (enc.east) -- (dec.west)
  node[lbl, midway, above] {memory $M$}
  node[lbl, midway, below] {cross-attn.};
\coordinate (lx) at ($(dec.east)+(0.3,0)$);
\draw[arr, dashed] (out.east) -- (out.east -| lx) -- (inY.east -| lx) -- (inY.east);

\end{tikzpicture}
\caption{GNN-Transformer architecture. A GCN embeds the ZX diagram and an embedding layer maps the source gate tokens to vectors. The two sequences are concatenated and encoded by a Pre-LN Transformer encoder. A Pre-LN decoder cross-attends to the encoder memory and generates the optimized circuit one token at a time (dashed: autoregressive feedback), with a qubit-count mask restricting outputs to valid gates.}
\label{fig:model}
\end{figure}

\Cref{fig:model} illustrates the model architecture: an encoder--decoder Transformer whose encoder processes two input modalities, analogous to vision--language models that pair a visual encoder with a language model~\cite{alayrac2022flamingo}, except that the second modality here is a graph.

In the graph branch, graph-convolutional (GCN) layers~\cite{kipf2017gcn}, each followed by a ReLU activation, map the node features $\bm{f}_v$ to the model dimension $d_{\mathrm{model}}$. Each layer adds self-loops and aggregates messages along the unweighted edges in the edge index (providing a topological prior while $\Csrc$ supplies exact gate and boundary-Hadamard semantics), combining each vertex's spider type, boundary indicator, and phase with features from neighboring spiders. Ordering vertices by PyZX index, the resulting node embeddings are packed into a sequence $H_D\in\R^{|V|\times d_{\mathrm{model}}}$.

The source tokens of length $S$ are embedded and scaled as $E=\sqrt{d_{\mathrm{model}}}\,\mathrm{Emb}(\Csrc)\allowbreak\in\R^{S\times d_{\mathrm{model}}}$, then concatenated with the graph embeddings along the sequence dimension:
\begin{equation}
X^{(0)}=\big[\,H_D\,;\,E\,\big]\in\R^{(|V|+S)\times d_{\mathrm{model}}},
\label{eq:concat}
\end{equation}
with a padding mask over the combined sequence. $X^{(0)}$ is processed by a stack of Pre-LN Transformer encoder layers~\cite{vaswani2017attention,xiong2020layer} followed by a final LayerNorm, producing the encoder memory $M$. Rotary position embeddings (RoPE)~\cite{su2024roformer} are applied in every self-attention layer across the concatenated sequence, allowing gate tokens and ZX spiders to attend to one another from the first layer onward.

The decoder is a stack of Pre-LN Transformer layers, each applying causal self-attention with RoPE, cross-attention to $M$, and a feed-forward network. A linear output head produces logits $z_t\in\R^{|\mathcal V|}$; encoder, decoder, and output embeddings are untied. Architectural hyperparameters (layer counts, attention heads, hidden dimensions, and total parameter count) are listed in \cref{tab:hp-model}. Supervised training uses teacher forcing on the shifted target sequence, $\pi_\theta(y_t\mid y_{<t},x)=\mathrm{softmax}(z_t)$.

\subsection{Decoding with validity masks}
\label{sec:decode}

Because the vocabulary covers up to $n_{\max}$ qubits, for an $n$-qubit circuit ($n < n_{\max}$) any token acting on a qubit index $q\ge n$ is invalid. For each $n$ we precompute the valid subset $\mathcal V_n\subset\mathcal V$ of tokens whose qubit indices are strictly less than $n$ (plus \texttt{<eos>}), and sample from the masked distribution
\begin{equation}
\pi_\theta(y_t\mid y_{<t},x)=\frac{\1[y_t\in\mathcal V_n]\,\exp(z_{t,y_t}/\mathcal{T})}{\sum_{v\in\mathcal V_n}\exp(z_{t,v}/\mathcal{T})}
\label{eq:mask}
\end{equation}
at temperature $\mathcal T$. Restricting support to $\mathcal V_n$ (e.g., 95 active gate tokens at $n=5$) guarantees that every sampled sequence parses into a syntactically valid $n$-qubit circuit. During RL rollouts we sample from the full valid support at $\mathcal T=1$ without top-$k$ or nucleus truncation.

\subsection{Verification by Haar-random state evolution}
\label{sec:verify}

Constructing dense $2^n\times 2^n$ unitaries is infeasible at 20 qubits, so the verifier instead compares circuits by evolving a random input state. For each input circuit, it draws a Haar-random state $\ket{\psi}\in\C^{2^n}$, evolves it through both the reference and each candidate circuit using the state-vector simulator Qulacs~\cite{suzuki2021qulacs}, and computes the state fidelity
\begin{equation}
F(\Cgen;\psi)=\big|\bra{\psi}U(\Cr)^{\dagger}U(\Cgen)\ket{\psi}\big|^2.
\label{eq:fidelity}
\end{equation}
A candidate is marked \emph{correct} if $|F-1|\le\tau_F$ with $\tau_F=10^{-4}$; unparseable sequences receive $F=0$. Because each reference $\Cr$ has already been verified against $\Csrc$ (\S\ref{sec:data}), equivalence to $\Cr$ implies equivalence to $\Csrc$. Taking the squared modulus in \cref{eq:fidelity} quotients out any global phase, matching \cref{eq:equiv}. Simulating an $m$-gate circuit requires $O(m\,2^n)$ time, compared with $O(4^n)$ memory merely to store a dense unitary.

This test is grounded in the relation between average state fidelity and process fidelity~\cite{horodecki1999general,nielsen2002simple} (Appendix~\ref{app:fidelity}). With $W=U(\Cr)^\dagger U(\Cgen)$ and $d=2^n$, the two circuits are equivalent up to global phase if and only if the process fidelity $F_{\mathrm p}=\allowbreak|\Tr W|^2/d^2$ equals one. For a Haar-random state $\ket{\psi}$, $F(\Cgen;\psi)$ is an unbiased estimator of the average state fidelity $\bar F=\allowbreak\E_\psi[F(\Cgen;\psi)]$, which satisfies
\begin{equation}
\bar F=\frac{d\,F_{\mathrm p}+1}{d+1},\qquad 1-\bar F=\frac{d}{d+1}\,(1-F_{\mathrm p}).
\label{eq:fid-mean}
\end{equation}
For $n\ge5$, the scaling factor $d/(d+1)$ lies in $[32/33,1)$, so expected single-state infidelity closely tracks process infidelity at the cost of one state-vector simulation. Although bounding worst-case false acceptance statistically requires many independent Haar-random states (Appendix~\ref{app:fidelity}), empirically we find that inequivalent candidates are consistently rejected already by a single Haar-random state: among the 32{,}000 candidates evaluated in \S\ref{sec:res-verify}, the first random state ($k=1$) rejected all 11{,}296 inequivalent candidates.

To guide RL post-training, the reward function ranks every correct candidate above every incorrect one, and among correct candidates it ranks by gate reduction relative to the reference (\cref{eq:delta}). For a candidate sequence $y$ decoding to $\Cgen$, let $c=\1[|F-1|\le\tau_F]$ and
\begin{equation}
g(\Cgen)=\clip\!\big(w_{\mathrm{all}}\Delta_{\mathrm{all}}+w_{\mathrm{2q}}\Delta_{\mathrm{2q}}+w_T\Delta_T,\,-1,\,1\big),
\label{eq:gatescore}
\end{equation}
with non-negative weights $(w_{\mathrm{all}},w_{\mathrm{2q}},w_T)$ emphasizing two-qubit and $T$ gates (Appendix~\ref{app:hyper}). The reward is
\begin{equation}
R(x,y;\psi)=\begin{cases}
b+g(\Cgen), & c=1,\\[2pt]
s\cdot\clip\!\big(F/(1-\tau_F),\,0,\,1\big), & c=0,
\end{cases}
\label{eq:reward}
\end{equation}
with $b=2$ (success bonus) and $s=0.2$ (failure scale). Correct candidates score in $[1,3]$ and incorrect ones in $[0,0.2]$. The condition $b-1>s$ guarantees the lexicographic order, and failures still receive a dense fidelity signal. The reward is non-negative, which MaxRL requires (\S\ref{sec:rl}).

\subsection{Supervised pretraining}
\label{sec:sft}

The policy is first trained by maximum likelihood on (source, diagram, reference) triples:
\begin{equation}
\mathcal L_{\mathrm{SFT}}(\theta)=-\E_{(x,y^{\mathrm{ref}})}\sum_{t=1}^{|y^{\mathrm{ref}}|}\log\pi_\theta\big(y^{\mathrm{ref}}_t\mid y^{\mathrm{ref}}_{<t},x\big),
\label{eq:sft}
\end{equation}
averaged over non-padding tokens. Supervised pretraining enables the model to learn valid circuit syntax, align gate tokens with ZX spiders, and approximate the PyZX optimization pipeline end-to-end. Because \cref{eq:sft} maximizes token-level likelihood against a single reference, however, it penalizes alternative equivalent circuits and provides no direct incentive for further gate reduction. Reinforcement learning fine-tuning overcomes both limitations.

\subsection{Reinforcement learning with group-relative advantages}
\label{sec:rl}

At each step we sample a batch of $B$ inputs $\{x_j\}_{j=1}^B$, draw a group of $G$ candidates $\{y_{j,i}\}_{i=1}^{G}\sim\pi_\theta(\cdot\mid x_j)$ per input on a shared random state $\ket{\psi_j}$, and evaluate them via the verifier and \cref{eq:reward} to obtain rewards $r_{j,i}=R(x_j,y_{j,i};\psi_j)$. A critic-free \emph{advantage estimator} maps each group of rewards to scalar weights $A_{j,i}$, and we perform one on-policy update minimizing
\begin{equation}
\mathcal{L}_{\mathrm{RL}}(\theta)=-\frac{1}{BG}\sum_{j=1}^{B}\sum_{i=1}^{G}A_{j,i}\sum_{t=1}^{|y_{j,i}|}\log\pi_\theta(y_{j,i,t}\mid y_{j,i,<t},x_j).
\label{eq:loss}
\end{equation}
We consider three advantage estimators (GRPO, MaxRL, and Hybrid), which compute $A_{j,i}$ from the group mean $\bar r_j=\frac1G\sum_{i=1}^G r_{j,i}$ and sample standard deviation $\sigma_j$:
\begin{align}
\textbf{GRPO:}\quad & A_{j,i}=\frac{r_{j,i}-\bar r_j}{\sigma_j+\varepsilon}\cdot\1[\sigma_j>\sigma_{\min}], \label{eq:adv-grpo}\\
\textbf{MaxRL:}\quad & A_{j,i}=\frac{r_{j,i}-\bar r_j}{\bar r_j+\varepsilon}\cdot\1[\bar r_j>0], \label{eq:adv-maxrl}\\
\textbf{Hybrid:}\quad & A_{j,i}=\begin{cases}A^{\mathrm{GRPO}}_{j,i}, & \text{group $j$ all correct},\\ A^{\mathrm{MaxRL}}_{j,i}, & \text{otherwise},\end{cases} \label{eq:adv-hybrid}
\end{align}
for small numerical guards $\varepsilon,\sigma_{\min}>0$. \Cref{alg:rl} summarizes the procedure.

\begin{algorithm}[t]
\caption{RL fine-tuning of \method{}}
\label{alg:rl}
\begin{algorithmic}[1]
\Require pretrained policy $\pi_\theta$, data mixture $\mathcal{D}$, group size $G$, estimator $\in\{$GRPO, MaxRL, Hybrid$\}$
\For{each step}
  \State sample inputs $\{x_j\}_{j=1}^{B}\sim\mathcal D$
  \For{each $j$}
    \State sample $y_{j,1..G}\sim\pi_\theta(\cdot\mid x_j)$ over $\mathcal V_{n_j}$ \Comment{\cref{eq:mask}}
    \State draw one Haar state $\ket{\psi_j}$; compute $F_{j,i}$, $c_{j,i}$, $r_{j,i}$ \Comment{\cref{eq:fidelity,eq:reward}}
    \State compute $A_{j,1..G}$ over the full group \Comment{\cref{eq:adv-grpo,eq:adv-maxrl,eq:adv-hybrid}}
  \EndFor
  \State one AdamW step on $\mathcal L_{\mathrm{RL}}$ (\cref{eq:loss}), gradient norm clipped to 1
\EndFor
\end{algorithmic}
\end{algorithm}

\emph{GRPO}~\cite{shao2024deepseekmath} normalizes centered rewards by the group standard deviation, making updates invariant to positive affine transformations of the reward. For a binary reward with pass probability $p=p_\theta(x,\psi)$, its expected gradient per input scales as $\nabla p/\sqrt{p(1-p)}=\nabla\,2\arcsin\!\sqrt{p}$, concentrating weight on inputs of intermediate difficulty while assigning weak updates when $p$ is small.

\emph{Continuous MaxRL} extends Maximum-Likelihood RL~\cite{tajwar2026maxrl}, which maximizes $\E\log p_\theta$ rather than $\E p_\theta$ so that hard inputs receive inverse-probability weight ($\propto 1/p$). For our graded reward, we replace $p_\theta$ with the expected reward $\mu_\theta(x,\psi)=\E_\theta[R(x,y;\psi)]$ while retaining \cref{eq:adv-maxrl}, estimating the gradient of $\E_{x,\psi}\log\mu_\theta(x,\psi)$ (Appendix~\ref{app:maxrl} proves this extension and its scale invariance and shift sensitivity). Under \cref{eq:reward}, for a fixed input, in an all-failing group with $\bar F>0$, the failure scale $s$ cancels to give $A_i\approx(F_i-\bar F)/\bar F$, driving the policy toward higher fidelity; in a mixed group on a hard input, rare correct candidates receive large positive advantages $(b+g-\bar r)/\bar r\gg 1$.

\emph{Hybrid} (\cref{eq:adv-hybrid}) addresses a limitation of MaxRL on already-solved inputs: once all $G$ candidates in a group are correct, rewards are $r_i=b+g_i$ and \cref{eq:adv-maxrl} reduces to $(g_i-\bar g)/(b+\bar g)$, where the bonus $b=2$ dilutes small differences in gate score ($\sigma_g\approx 0.1$) by roughly $20\times$ compared with GRPO. Hybrid therefore applies MaxRL whenever a group contains any incorrect candidate and switches to GRPO when all $G$ candidates are correct (Appendix~\ref{app:hybrid}): in an all-correct group, $b$ cancels out of $A_i=(g_i-\bar g)/(\sigma_g+\varepsilon)$, leaving an unattenuated advantage that directly optimizes gate count among valid candidates. Because all $G$ candidates pass with probability $p^G$, this switch rarely triggers on hard inputs ($p^G\ll 1$, retaining MaxRL) and smoothly takes over as accuracy approaches one, without extra hyperparameters.

\section{Training Data}
\label{sec:data}

\method{} learns from triples $(\Csrc, D, \Cr)$: a random source circuit, a simplified ZX diagram of it, and an equivalent reference circuit extracted from that diagram. All triples are generated synthetically, so the corpus can grow to millions of examples and every reference is checked for equivalence before it is stored. \Cref{fig:pipeline}(1) summarizes the generator.

\subsection{Random source circuits}
\label{sec:data-random}

Each sample starts with a qubit count $n$ and a gate count $m$, uniformly sampled from $\{5,\dots,20\}$ and $\{m_{\min},\dots,m_{\max}\}$ (\S\ref{sec:data-buckets}), respectively. The circuit is then built gate by gate. Each gate type is drawn independently from the fixed distribution in \cref{tab:gate-probs} and placed on a uniformly random qubit (single-qubit gates) or a uniformly random ordered pair of distinct qubits (CNOT, CZ). A phase gate $R_Z(\ell\pi/4)$ takes $\ell$ uniformly from $\{1,2,3,5,6,7\}$.

\begin{table}[t]
\centering
\small
\caption{Gate distribution of the random source circuits.}
\label{tab:gate-probs}
\begin{tabular}{@{}lcccccccc@{}}
\toprule
gate & $H$ & $X$ & $Z$ & $S$ & $T$ & $R_Z(\ell\pi/4)$ & CNOT & CZ \\
\midrule
probability & 0.1 & 0.1 & 0.1 & 0.1 & 0.2 & 0.1 & 0.2 & 0.1 \\
\bottomrule
\end{tabular}
\end{table}

\subsection{ZX simplification and reference extraction}
\label{sec:data-zx}

The reference is produced with PyZX~\cite{kissinger2020pyzx} in five steps:
\begin{enumerate}
  \item The source is converted to a ZX diagram $D_0$.
  \item Phase teleportation~\cite{kissinger2020reducing} moves non-Clifford phases so that they can fuse without altering the graph structure, giving $D_1$.
  \item $D_1$ is converted to graph-like form~\cite{duncan2020graph} and simplified with the causal-flow-preserving heuristic of Holker~\cite{holker2023causal} to reduce two-qubit gates while keeping the diagram directly extractable~\cite{backens2021there}, giving $D_2$ (used as the input diagram $D$ in \S\ref{sec:method}).
  \item A circuit is extracted from $D_2$ and decomposed into basic gates.
  \item It is post-processed by PyZX's gate-level optimization, with and without SWAP-aware rules. Of the two results, the one with fewer two-qubit gates becomes $\Cr$.
\end{enumerate}

Every extracted reference $\Cr$ is verified against $\Csrc$ using the Haar-state test of \S\ref{sec:verify} with tolerance $10^{-6}$ (discarding and regenerating any sample that fails, to guard against extraction bugs). PyZX expresses extracted phases in the $R_Z(\ell\pi/4)$ basis (with $X$-axis rotations conjugated by $H$), so every token in $\Csrc$ and $\Cr$ belongs to the vocabulary $\mathcal V$ of \S\ref{sec:repr}. We retain $\Cr$ even when it contains more gates than $\Csrc$: supervised pretraining imitates the PyZX output directly, while subsequent RL fine-tuning rewards further gate reduction. Conditioning on $D_2$ alongside $\Csrc$ provides the model with both the explicit input gate sequence and the simplified ZX graph structure prior to circuit extraction.

\subsection{Size buckets}
\label{sec:data-buckets}

For curriculum training (\S\ref{sec:setup}), the corpus is partitioned by the gate count of the \emph{source} circuit into four buckets: $[20,100]$, $[101,200]$, $[201,300]$, and $[301,400]$ gates. Each bucket maintains independent training, validation, and test splits (\cref{tab:data-sizes}). Supervised pretraining uses the $[20,100]$ bucket, the two-stage RL curriculum (\S\ref{sec:setup}) progressively incorporates $[101,200]$ and $[201,300]$, and $[301,400]$ is held out to evaluate length generalization to unseen longer circuits. For benchmark evaluation (\S\ref{sec:results}), we draw a fixed subset of 9{,}000 circuits from each bucket's test split, along with a qubit-stratified subset of 992 circuits (62 per qubit count across $n\in\{5,\dots,20\}$) for high-budget ($K\le 64$) sampling experiments.

\begin{table}[t]
\centering
\small
\caption{Data buckets (5--20 qubits): number of samples in each split.}
\label{tab:data-sizes}
\begin{tabular}{@{}lrrr@{}}
\toprule
source gates & train & valid & test \\
\midrule
20--100   & 7{,}000{,}000 & 50{,}000 & 100{,}000 \\
101--200  & 2{,}013{,}419 & 10{,}040 & 9{,}959 \\
201--300  & 2{,}013{,}082 & 9{,}978  & 10{,}033 \\
301--400  & 2{,}012{,}564 & 9{,}898  & 9{,}940 \\
\bottomrule
\end{tabular}
\end{table}

\section{Experimental Setup and Training Pipeline}
\label{sec:setup}

Training proceeds in three stages (\cref{fig:stages}). Hyperparameters and software/hardware details are listed in Appendix~\ref{app:hyper}.

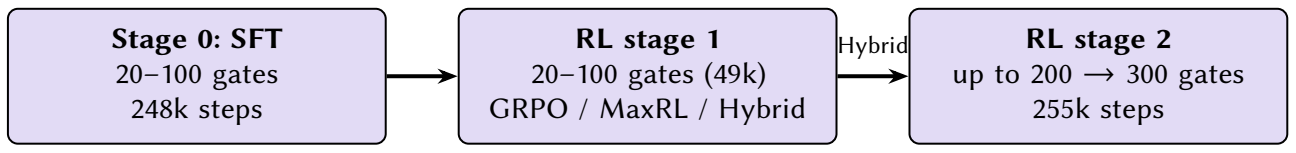
\begin{figure}[t]
\centering
\resizebox{\columnwidth}{!}{%
\begin{tikzpicture}[font=\sffamily\scriptsize,
  st/.style={draw, rounded corners=2pt, fill=newcomp, align=center, minimum height=0.95cm, text width=2.45cm},
  arr/.style={-{Stealth[length=4pt]}, thick}]
\node[st] (s0) {\textbf{Stage 0: SFT}\\ 20--100 gates\\ 248k steps};
\node[st, right=0.50cm of s0] (s1) {\textbf{RL stage 1}\\ 20--100 gates (49k)\\ GRPO / MaxRL / Hybrid};
\node[st, right=0.50cm of s1] (s2) {\textbf{RL stage 2}\\ up to 200 $\to$ 300 gates\\ 255k steps};
\draw[arr] (s0) -- (s1);
\draw[arr] (s1) -- node[above, font=\sffamily\tiny] {Hybrid} (s2);
\end{tikzpicture}}
\caption{Training pipeline. Stage 1 compares advantage estimators at fixed circuit size. Stage 2 initializes from the stage-1 Hybrid policy and progressively introduces larger circuits.}
\label{fig:stages}
\end{figure}

\emph{Stage 0: supervised pretraining.} The 174M-parameter model of \S\ref{sec:model} is pretrained with \cref{eq:sft} to reproduce the PyZX reference on the 20--100-gate bucket conditioned on $D=D_2$, using AdamW~\cite{loshchilov2019decoupled} with a global batch size of 256 (\cref{tab:hp-train}). Validation cross-entropy reaches $0.1303$ (perplexity $1.14$) at approximately 248{,}000 steps ($\approx 9$ epochs), which initializes all RL runs.

\emph{Stage 1: comparing advantage estimators.} Stage 1 fine-tunes the pretrained policy on 20--100-gate circuits to isolate the effect of the advantage estimator while holding all other settings fixed across GRPO, MaxRL, and Hybrid: identical initialization, training data, and reward (\cref{eq:reward}); identical on-policy objective (\cref{eq:loss}) with unit gradient-norm clipping; identical rollout sampling ($G=32$, $\mathcal T=1$, full valid support); a constant learning rate of $10^{-5}$; and identical rollout and validation seeds. Each step samples 32 inputs (1{,}024 candidates per step); checkpoints are selected by mean validation reward on 20--100-gate validation circuits, and we evaluate step 49{,}152 of each run.

\emph{Stage 2: curriculum over circuit size.} Because a policy trained solely on 20--100-gate circuits generalizes poorly to 200- or 300-gate inputs~\cite{saeedi2026close}, we initialize from the stage-1 Hybrid policy (step 49{,}152) and apply a two-stage size curriculum~\cite{bengio2009curriculum} using weighted mixtures of the size buckets (\S\ref{sec:data-buckets}), assigning $70\%$ weight to the newly introduced bucket while replaying shorter buckets to prevent forgetting:
\begin{enumerate}
  \item \emph{Stage 2a} (up to 200 gates, steps 0--116{,}736): 20--100 : 101--200 $=0.30:0.70$;
  \item \emph{Stage 2b} (up to 300 gates, from step 116{,}736; evaluated at step 254{,}976): 20--100 : 101--200 : 201--300 $=0.15:0.15:0.70$.
\end{enumerate}
Each step samples a global batch of 64 inputs with $G=32$ rollouts per input (2{,}048 candidates per step), selecting checkpoints by mean validation reward across a fixed multi-bucket validation pool. The 301--400-gate bucket is never used in training and serves strictly to evaluate out-of-distribution length generalization.

\section{Experimental Results}
\label{sec:results}

We evaluate all models on the held-out sets of 9{,}000 random circuits per size bucket (\S\ref{sec:data-buckets}).

\subsection{Metrics and verification reliability}
\label{sec:res-metrics}
\label{sec:res-verify}

\emph{Circuit accuracy} (pass@$K$) is the fraction of test circuits with at least one valid candidate among $K$ samples~\cite{chen2021evaluating}, with pass@$1$ estimated via $c/K$ where $c$ is the number of valid samples. Among valid candidates for a given input, we select the lexicographic minimum under $(\Nall, \Ntq, \NT)$ (ranking by $\kappa(\Cgen)$ or $(\Ntq,\NT,\Nall)$ gives virtually identical results at $\mathcal T=0.2$) and report gate reductions $N_k(\Csrc)-N_k(\Cgen)$ alongside PyZX's $N_k(\Csrc)-N_k(\Cr)$. The \emph{better-than-reference} (and \emph{worse-than-reference}) rate is the fraction of a model's solved circuits on which its selected circuit uses strictly fewer (or strictly more) gates than $\Cr$. To avoid selection bias across models with different solve rates, mean gate reductions are always computed on the \emph{common success set} solved by all compared models in a table or plot.

\begin{figure}[t]
\centering
\includegraphics[width=\columnwidth]{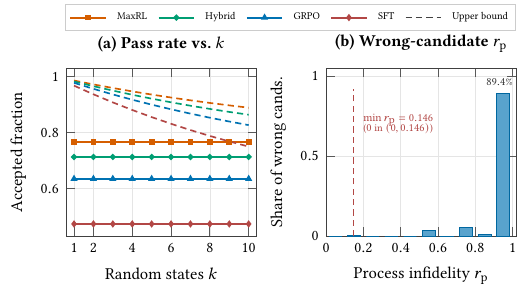}
\caption{Validation of random-state verification on 32{,}000 candidates ($5\le n\le 14$). \textbf{(a)} Fraction passing $k$ Haar-random state tests (solid) versus the upper bound $f_c+(1-f_c)(15/16)^k$ (dashed); empirical acceptance equals the exact equivalence rate $f_c$ already at $k=1$. \textbf{(b)} Exact process infidelity $r_{\mathrm p}=1-F_{\mathrm p}$ across all 11{,}296 inequivalent candidates: none lie in $(10^{-8},0.146)$.}
\label{fig:verify-c5}
\end{figure}

\begin{figure*}[t]
\centering
\includegraphics[width=\textwidth]{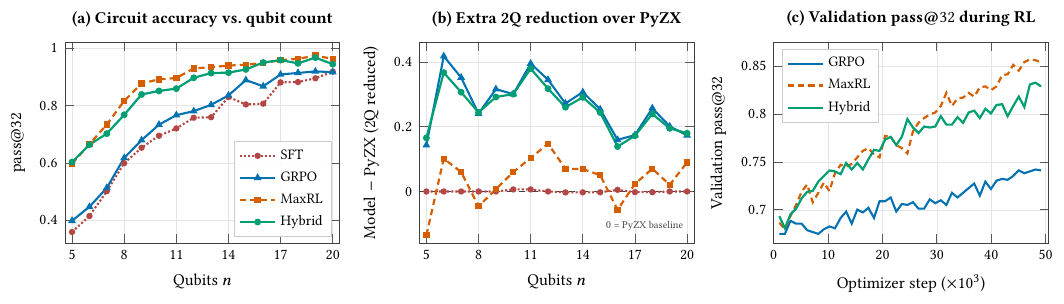}
\caption{Stage-1 advantage-estimator comparison on 20--100-gate circuits (step 49{,}152). \textbf{(a)} Test pass@$32$ by qubit count. \textbf{(b)} Extra two-qubit gate reduction over PyZX on the common success set ($n_{\mathrm{com}}=5{,}443$). \textbf{(c)} Validation pass@$32$ during stage-1 training.}
\label{fig:estimators}
\end{figure*}

Before presenting benchmark comparisons, we validate our single-state verifier ($k=1$, $\tau_F=10^{-4}$) in two ways. First, on 32{,}000 candidates sampled across SFT, GRPO, MaxRL, and Hybrid on 20--100-gate circuits ($5\le n\le 14$), we compare $k=1,\dots,10$ independent Haar-random state tests against exact process infidelity $r_{\mathrm p}=1-F_{\mathrm p}$ (\cref{fig:verify-c5}). Already at $k=1$, the random-state test rejects all $11{,}296$ inequivalent candidates ($0$ false acceptances): zero inequivalent candidates fall in $(10^{-8},0.146)$, matching the exact single-gate lower bound $r_{\mathrm p,\min}^{(1)}=\sin^2(\pi/8)\approx 0.1464$ (Appendix~\ref{app:fidelity}), and $89.4\%$ have $r_{\mathrm p}\ge 0.9$.

Second, we subject all $80{,}674$ selected best candidates across every model--bucket evaluation reported below to deterministic certification against $\Cr$ via symbolic ZX simplification of $\Cr^\dagger\Cgen$~\cite{peham2022equivalence}, supplemented by exact unitary comparison for all $24{,}910$ candidates with $n\le 10$ (Appendix~\ref{app:fidelity}). Every single candidate passes ($100\%$ certified, zero rejections).

\subsection{Advantage estimators: GRPO, MaxRL, and Hybrid}
\label{sec:res-estimators}

\begin{table}[t]
\centering
\small
\setlength{\tabcolsep}{3.6pt}
\caption{Controlled stage-1 comparison on 9{,}000 held-out 20--100-gate circuits ($K=32$, $\mathcal T=0.2$, step 49{,}152). Gate reductions are on the common success set ($n_{\mathrm{com}}=5{,}443$); \emph{better} and \emph{worse} (2Q) are relative to PyZX on each model's solved set.}
\label{tab:estimator-results}
\begin{tabular}{@{}lccccccc@{}}
\toprule
 & & & \multicolumn{3}{c}{gates reduced} & \multicolumn{2}{c}{vs.\ ref.\ (2Q)} \\
\cmidrule(lr){4-6}\cmidrule(l){7-8}
System & pass@$1$ & pass@$32$ & all & 2Q & $T$ & better & worse \\
\midrule
PyZX ref. & -- & -- & 14.19 & 0.79 & \textbf{6.21} & -- & -- \\
SFT (init.) & 60.8\% & 71.6\% & 14.24 & 0.79 & \textbf{6.21} & 0.1\% & \textbf{0.3\%} \\
GRPO   & 72.7\% & 74.7\% & \textbf{15.61} & \textbf{1.05} & 6.20 & 21.1\% & 10.6\% \\
MaxRL  & \textbf{83.4\%} & \textbf{87.6\%} & 14.88 & 0.84 & 6.19 & 20.4\% & 27.9\% \\
Hybrid & 80.4\% & 85.5\% & 15.58 & 1.04 & 6.20 & \textbf{21.9\%} & 16.0\% \\
\bottomrule
\end{tabular}
\end{table}

\begin{table*}[t]
\centering
\small
\caption{Main results on held-out random circuits (9{,}000 circuits per bucket, 5--20 qubits, $K=8$, $\mathcal T=0.2$). \emph{S2a-Hybrid} (step 116{,}736) is trained up to 200 gates and \emph{S2b-Hybrid} (step 254{,}976) up to 300 gates (omitting policies with $0\%$ pass@$8$); the unseen 301--400-gate bucket evaluates length generalization. Mean gate reductions are relative to $\Csrc$ on each trained bucket's common success set ($n_{\mathrm{com}}=5{,}466$, $1{,}128$, and $797$) and on the $690$ solved circuits for 301--400. \emph{Better (all)} and \emph{better (2Q)} are fractions of each model's own solved circuits strictly beating $\Cr$.}
\label{tab:main}
\begin{tabular}{@{}llccccccc@{}}
\toprule
 & & & & \multicolumn{3}{c}{gates reduced (common set)} & \multicolumn{2}{c}{better than ref.} \\
\cmidrule(lr){5-7}\cmidrule(l){8-9}
Bucket & System & pass@$1$ & pass@$8$ & all & 2Q & $T$ & all & 2Q \\
\midrule
\multirow{5}{*}{20--100}
 & PyZX reference & -- & -- & 14.21 & 0.82 & \textbf{6.21} & -- & -- \\
 & \method{} (SFT) & 60.8\% & 67.7\% & 14.25 & 0.82 & \textbf{6.21} & 5.0\% & 0.1\% \\
 & \method{} (S1-Hybrid) & 80.4\% & 83.5\% & 15.57 & 1.06 & 6.20 & 59.5\% & 21.8\% \\
 & \method{} (S2a-Hybrid) & 90.8\% & 92.9\% & \textbf{15.69} & \textbf{1.07} & 6.20 & \textbf{60.2\%} & \textbf{23.7\%} \\
 & \method{} (S2b-Hybrid) & \textbf{92.0\%} & \textbf{94.1\%} & 15.67 & 1.06 & 6.20 & 59.7\% & 23.6\% \\
\midrule
\multirow{4}{*}{101--200}
 & PyZX reference & -- & -- & 44.57 & 2.31 & \textbf{17.76} & -- & -- \\
 & \method{} (S1-Hybrid) & 12.5\% & 15.4\% & \textbf{47.06} & 2.39 & 17.65 & \textbf{69.3\%} & \textbf{35.1\%} \\
 & \method{} (S2a-Hybrid) & 60.7\% & 68.7\% & 46.87 & \textbf{2.43} & 17.65 & 41.8\% & 29.9\% \\
 & \method{} (S2b-Hybrid) & \textbf{71.4\%} & \textbf{78.8\%} & 46.75 & \textbf{2.43} & 17.65 & 38.5\% & 28.6\% \\
\midrule
\multirow{3}{*}{201--300}
 & PyZX reference & -- & -- & \textbf{93.79} & \textbf{5.13} & \textbf{34.56} & -- & -- \\
 & \method{} (S2a-Hybrid) & 6.8\% & 10.2\% & 88.20 & 3.43 & 34.26 & \textbf{19.1\%} & \textbf{25.0\%} \\
 & \method{} (S2b-Hybrid) & \textbf{38.8\%} & \textbf{50.7\%} & 87.90 & 3.34 & 34.25 & 9.2\% & 20.2\% \\
\midrule
301--400 (unseen)
 & \method{} (S2b-Hybrid) & 4.1\% & 7.7\% & 129.47 & 4.50 & 51.75 & -- & -- \\
\bottomrule
\end{tabular}
\end{table*}

\begin{figure*}[t]
\centering
\includegraphics[width=\textwidth]{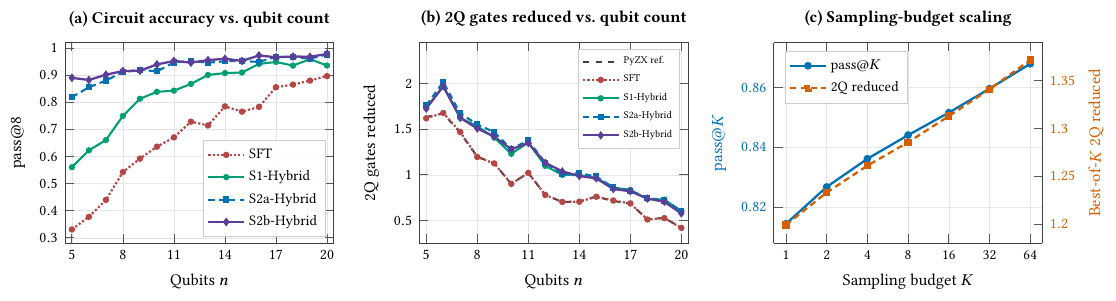}
\caption{Main evaluation on 20--100-gate circuits (5--20 qubits). \textbf{(a)} Test pass@$8$ by qubit count. \textbf{(b)} Mean two-qubit gates reduced on the common success set ($n_{\mathrm{com}}=5{,}466$). \textbf{(c)} Inference-time scaling of S1-Hybrid ($K\in\{1,\dots,64\}$ on 992 stratified circuits; best-of-$K$ 2Q reduction assigns zero reduction when no sample in $K$ is valid).}
\label{fig:main-eval}
\end{figure*}

\begin{figure*}[t]
\centering
\includegraphics[width=\textwidth]{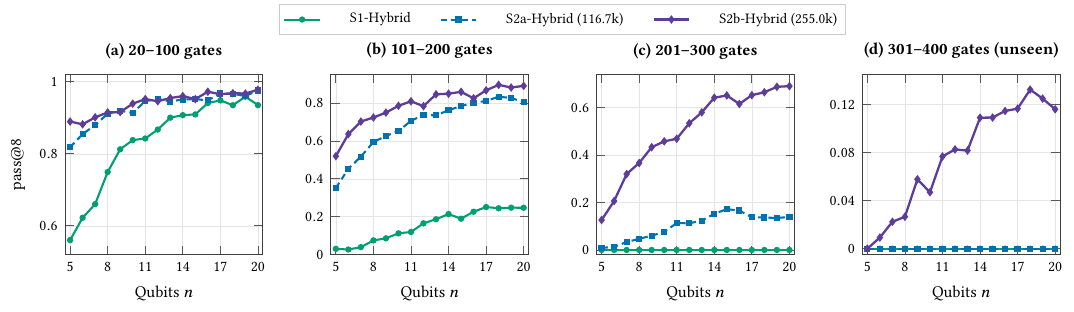}
\caption{Per-qubit pass@$8$ of S1-Hybrid, S2a-Hybrid (trained up to 200 gates), and S2b-Hybrid (trained up to 300 gates) on \textbf{(a)} 20--100, \textbf{(b)} 101--200, \textbf{(c)} 201--300, and \textbf{(d)} unseen 301--400 gates.}
\label{fig:curriculum}
\end{figure*}

\Cref{tab:estimator-results} and \cref{fig:estimators} first compare the three stage-1 estimators under identical initialization and compute (\S\ref{sec:setup}), confirming the analysis of \S\ref{sec:rl}:
\begin{enumerate}
  \item \emph{Correctness on hard inputs.} MaxRL ($87.6\%$ pass@$32$) and Hybrid ($85.5\%$) far outperform GRPO ($74.7\%$, only $+3.1$ pp over SFT), with the largest gap on dense small-qubit circuits ($59.8\%$ and $60.3\%$ vs.\ $39.9\%$ at $n=5$; \cref{fig:estimators}(a)), where MaxRL's inverse-mean weighting prioritizes low-pass-rate inputs.

  \item \emph{Gate-cost optimization on solved inputs.} MaxRL yields much smaller gate reductions than GRPO ($0.84$ vs.\ $1.05$ 2Q; $27.9\%$ worse than PyZX) because the success bonus $b$ dilutes gate-cost differences in all-correct groups. Switching to GRPO normalization on all-correct groups allows Hybrid to match GRPO's gate reduction ($1.04$ 2Q, $21.9\%$ better than PyZX; \cref{fig:estimators}(b)) while staying within $2.1$ pp of MaxRL's pass@$32$.

  \item \emph{Diversity and training dynamics.} The pass@$32-$pass@$1$ gap is only $2.0$ pp for GRPO versus $4.2$ pp for MaxRL and $5.1$ pp for Hybrid, indicating narrower mode collapse under GRPO, while MaxRL and Hybrid improve validation accuracy steadily throughout training (\cref{fig:estimators}(c); Appendix~\ref{app:results}).
\end{enumerate}
Hybrid thus provides the best trade-off between solve rate and gate reduction, motivating its selection as the base policy for Stage 2.

\subsection{Main results}
\label{sec:res-main}

\Cref{tab:main} and \cref{fig:main-eval,fig:curriculum} summarize three central findings across the four size buckets.

First, verifier-guided RL moves the model from imitating PyZX to surpassing it in both reliability and gate count. Under supervised pretraining, the model largely copies its teacher, matching PyZX's two-qubit gate count on nearly all solved circuits. RL post-training raises pass@$8$ on 20--100-gate circuits from $67.7\%$ to $94.1\%$, with the largest gains on dense, small-qubit circuits where extraction is hardest (\cref{fig:main-eval}(a)), and consistently reduces more total and two-qubit gates than PyZX across all qubit counts (\cref{fig:main-eval}(b)). Across the three trained buckets in \cref{tab:main}, the final curriculum policy (S2b-Hybrid) finds strictly fewer two-qubit gates than PyZX on $20\%$--$29\%$ of solved circuits while closely matching PyZX in $T$ count (\S\ref{sec:discussion}); selecting the circuit with fewer two-qubit gates between $\Cr$ and $\Cgen$ yields net full-test-set reductions of $+0.47$, $+0.51$, and $+0.26$ two-qubit gates per circuit over PyZX alone.

Second, the two-stage size curriculum overcomes the sharp length degradation of fixed-size RL while inducing positive backward transfer (\cref{fig:curriculum}). Whereas S1-Hybrid trained solely on short circuits drops to $15.4\%$ pass@$8$ on 101--200 gates and $0\%$ beyond, progressive curriculum training scales pass@$8$ to $68.7\%$ (S2a-Hybrid) and $78.8\%$ (S2b-Hybrid) on 101--200 gates, reaches $50.7\%$ on 201--300 gates, and achieves $7.7\%$ pass@$8$ on unseen 301--400-gate circuits (\cref{fig:curriculum}(b--d); $0\%$ length-cap truncation or parse failures, with errors arising solely from inequivalence). Replaying shorter buckets not only prevents forgetting, but further improves pass@$8$ on 20--100 gates ($83.5\%\to 92.9\%\to 94.1\%$) and 101--200 gates ($68.7\%\to 78.8\%$) across Stages 2a and 2b (\cref{fig:curriculum}(a,b); validation trajectories in \cref{fig:app-s2}(a,b)) while preserving gate-reduction quality (\cref{tab:main}).

Third, increasing the test-time sampling budget from $K=1$ to $64$ steadily improves both pass@$K$ and best-of-$K$ two-qubit gate reduction without retraining (\cref{fig:main-eval}(c)), with especially pronounced gains on longer circuits (e.g., lifting pass@$K$ from $38.5\%$ to $56.3\%$ on 201--300 gates; \cref{fig:app-s2}(c)).

\section{Discussion and Conclusion}
\label{sec:conclusion}
\label{sec:discussion}

We formulated quantum circuit optimization as graph-conditioned sequence-to-sequence generation. \method{} jointly encodes a source circuit's gate-token sequence and simplified ZX diagram with a GNN-Transformer, and is pretrained on PyZX-optimized references before reinforcement learning post-training with a Haar-random state verifier and a gate-reduction reward. Extending MaxRL to continuous non-negative rewards and introducing the Hybrid advantage estimator combines MaxRL's high solve rate on hard circuits with GRPO's unattenuated gate-count optimization on solved circuits. Together with a two-stage size curriculum and symbolic ZX certification of all selected benchmark solutions, verifier-guided RL moves \method{} well beyond supervised imitation in both equivalence reliability and two-qubit gate reduction.

On the 20--100- and 101--200-gate benchmarks, RL consistently reduces more two-qubit gates on average than PyZX while closely matching PyZX's $T$-gate counts (within $0.01$--$0.31$ gates; fault-tolerant settings can further prioritize $T$ count via a larger $w_T$). This contrast reflects the reference pipeline (\S\ref{sec:data-zx}): phase teleportation~\cite{kissinger2020reducing} already fuses essentially all collapsible non-Clifford phases on random Clifford+$T$ circuits, leaving little headroom in $T$ count, whereas causal-flow extraction introduces auxiliary two-qubit gates~\cite{staudacher2023reducing} that graph-conditioned autoregressive decoding learns to avoid. On the 201--300-gate bucket, mean gate reduction still trails PyZX because moderate single-sample accuracy ($p$) makes all-correct rollout groups rare ($p^G\ll 1$), keeping Hybrid primarily in its MaxRL correctness regime; following Hybrid training with a subsequent GRPO stage once accuracy is established offers a direct remedy.

For circuit scale, fixed-size policies fail on deeper inputs, especially at high per-qubit gate density $m/n$ (small $n$, \cref{fig:curriculum}), due to compounding autoregressive errors rather than decoding length caps~\cite{saeedi2026close}. Our size curriculum addresses this progressively: although our experiments stop training at 201--300 source gates (already generalizing to unseen 301--400-gate circuits), the curriculum itself can continue advancing to longer buckets to handle deeper circuits, complemented at inference time by larger sampling budgets ($K>1$) without retraining. Extending training and evaluation from synthetic random circuits and a single causal-flow ZX baseline to structured algorithmic benchmarks is a natural next step.

Finally, two directions will broaden practical deployment. First, state-vector verification scales exponentially with qubit count (reaching hundreds of milliseconds per candidate at $n=20$), limiting online RL training to roughly 20 qubits; scaling to wider registers motivates fast symbolic ZX rewriting or approximate tensor-network verifiers. Second, because hardware constraints enter our formulation only through the qubit-indexed gate vocabulary, validity mask, and gate-cost reward, adapting these components offers a direct path to larger qubit registers, native device gate sets, and connectivity-aware routing.

\section*{Code Availability}
The inference and equivalence-verification code for \method{}, together with the trained S2b-Hybrid model checkpoint, is publicly available at \url{https://github.com/Xiao-diLi/GraftQ}.

\begin{acks}
We thank Xue Chen, Xiaotian Nie, Hui Zhai, and Si Wei for helpful discussions.
\end{acks}

\bibliographystyle{ACM-Reference-Format-unsrt}
\bibliography{references}

\appendix
\section{Hyperparameters and Implementation Details}
\label{app:hyper}

\Cref{tab:hp-model} lists the model architecture and \cref{tab:hp-train} lists the training hyperparameters for all three stages.

\paragraph{Software and hardware.} Our implementation uses PyTorch~\cite{paszke2019pytorch} and PyTorch Lightning, PyTorch Geometric~\cite{fey2019fast} for GCN layers, a customized PyZX~\cite{kissinger2020pyzx} incorporating the causal-flow-preserving simplification of Holker~\cite{holker2023causal}, and Qulacs~\cite{suzuki2021qulacs} for state-vector simulation. Supervised pretraining runs on four NVIDIA A100 (48\,GB) GPUs; each stage-1 RL run uses a single A100 (48\,GB) GPU ($\approx 85$ GPU-hours); and stage-2 curriculum runs use four or eight NVIDIA A100 (48\,GB) / A800 (80\,GB) GPUs with a fixed global batch size of 64 inputs per step.

\begin{table}[h]
\centering\small
\caption{Model architecture hyperparameters.}
\label{tab:hp-model}
\begin{tabular}{@{}ll@{}}
\toprule
parameter & value \\
\midrule
$d_{\mathrm{model}}$ & 1024 \\
encoder layers & 8 \\
decoder layers & 8 \\
attention heads & 16 \\
head dimension $d_k$ & 64 \\
FFN width & 2048 \\
FFN activation & ReLU \\
normalization position & Pre-LN \\
position encoding & RoPE (self-attention only) \\
dropout & 0 \\
GCN layers & 4 \\
GCN in-features & $2n_{\max}+5=45$ \\
GCN intermediate / out & 1024 \\
vocab size $|\mathcal{V}|$ & 984 \\
max qubits $n_{\max}$ & 20 \\
total parameters & ${\approx}174\text{M}$ \\
\bottomrule
\end{tabular}
\end{table}

\begin{table}[h]
\centering\small
\setlength{\tabcolsep}{3.6pt}
\caption{Training hyperparameters.}
\label{tab:hp-train}
\begin{tabular}{@{}llll@{}}
\toprule
parameter & SFT & RL stage 1 & RL stage 2 \\
\midrule
optimizer & AdamW & AdamW & AdamW \\
$(\beta_1,\beta_2)$ & $(0.9,0.98)$ & $(0.9,0.98)$ & $(0.9,0.98)$ \\
$\epsilon$ & $10^{-9}$ & $10^{-9}$ & $10^{-9}$ \\
weight decay & 0.01 & 0 & 0 \\
peak LR & $2\!\times\!10^{-4}$ & $10^{-5}$ & $10^{-5}$ \\
LR schedule & warmup+cosine & constant & constant \\
warmup steps & 3000 & -- & -- \\
gradient clip & 1.0 & 1.0 & 1.0 \\
global batch $B$ & 256 & 32 & 64 \\
group size $G$ & -- & 32 & 32 \\
updates / batch & -- & 1 (on-policy) & 1 (on-policy) \\
rollout $\mathcal T$ & -- & 1.0 & 1.0 \\
valid $\mathcal T$ & -- & 0.1 & 0.1 \\
valid interval & 2000 & 1024 & 1024 \\
valid samples & -- & 1024 & 500 / bucket \\
steps (eval) & ${\approx}248$k (${\approx}9$ ep.) & 49{,}152 & 116.7k / 255.0k \\
precision & bf16 & bf16 & bf16 \\
KL penalty & -- & none & none \\
fidelity tol.\ $\tau_F$ & -- & $10^{-4}$ & $10^{-4}$ \\
\bottomrule
\end{tabular}
\end{table}

\paragraph{Reward, advantage, and decoding parameters.} Reward (\cref{eq:reward}): $b=2$, $s=0.2$, $(w_{\mathrm{all}},w_{\mathrm{2q}},w_T)=(1,2,2)$, $\tau_F=10^{-4}$ (with a $10^{-6}$ denominator guard in \cref{eq:delta} when $N_k(\Cr)=0$; assigning a larger weight $w_T$ does not push $T$-gate reduction past PyZX, because PyZX's phase teleportation already reduces $T$ count very effectively on these circuits; see \S\ref{sec:discussion}); numerical guards (\cref{eq:adv-grpo,eq:adv-maxrl}): $\varepsilon=10^{-8}$, $\sigma_{\min}=10^{-4}$. Test decoding uses temperature $\mathcal T=0.2$ without top-$k$ or top-$p$ truncation; RL rollouts and test decoding cap output length at $1.2\times$ the longest reference in the batch, while training-time validation uses a 512-token cap. Each RL rollout batch performs a single on-policy optimizer step without a KL penalty (\cref{eq:loss}).

\section{MaxRL with Continuous Rewards}
\label{app:maxrl}

We extend Maximum-Likelihood Reinforcement Learning~\cite{tajwar2026maxrl} to continuous non-negative rewards and prove three results (\cref{prop:cont,prop:lift,prop:shift}) used in \S\ref{sec:rl}.

Throughout this appendix, fix an input $x$ and a shared Haar-random verifier state $\ket{\psi}$ (\cref{alg:rl}); the policy $\pi_\theta(\cdot\mid x)$ observes only $x$, so its score function $S=\nabla_\theta\log\pi_\theta(y\mid x)$ is independent of $\psi$. Let $R=R(x,y;\psi)\ge0$ denote the single-state reward (\cref{eq:reward}) with conditional expectation $\mu_\theta(x,\psi)=\E_\theta[R(x,y;\psi)]>0$, where $\E_\theta[\cdot]$ denotes expectation under $y\sim\pi_\theta(\cdot\mid x)$ for fixed $(x,\psi)$.

\subsection{Continuous MaxRL gradient}

\begin{proposition}
\label{prop:cont}
The gradient of $J_{\mathrm{MaxRL}}(\theta)=\E_{x,\psi}\log\mu_\theta(x,\psi)$ with respect to $\theta$ satisfies
\begin{equation}
\nabla_\theta J_{\mathrm{MaxRL}} = \E_{x,\psi}\!\left[\E_\theta\!\left[\frac{R}{\mu_\theta(x,\psi)}\,S\right]\right]
= \E_{x,\psi}\!\left[\E_q[S]\right],
\end{equation}
where $q(y\mid x,\psi)\propto\pi_\theta(y\mid x)\,R(x,y;\psi)$ is the reward-tilted distribution.
Conditioned on $(x,\psi)$, the $G$ group rewards $\{r_i\}_{i=1}^G$ are i.i.d.\ with mean $\mu_\theta(x,\psi)$; replacing $\mu_\theta(x,\psi)$ with the group sample mean $\bar r=\frac{1}{G}\sum_{i=1}^G r_i$ yields the self-normalized advantage $A_i=(r_i-\bar r)/\bar r$, for which $\hat g=(BG)^{-1}\sum_{j,i}A_{j,i}S_{j,i}$ is an asymptotically unbiased estimator of $\nabla_\theta J_{\mathrm{MaxRL}}$ with finite-group bias $O(1/G)$.
\end{proposition}
\begin{proof}
For fixed $(x,\psi)$, $R(x,y;\psi)$ does not depend on $\theta$, so the log-derivative identity gives $\nabla_\theta\log\mu_\theta(x,\psi)=\E_\theta[RS]/\mu_\theta(x,\psi)=\E_\theta[(R/\mu_\theta)S]$. The reward-tilted rewrite follows from $q(y\mid x,\psi)=\pi_\theta(y\mid x)R(x,y;\psi)/\mu_\theta(x,\psi)$. Since $\E_\theta[S]=0$, $\E_\theta[(R-\mu_\theta)S/\mu_\theta]=\E_\theta[RS]/\mu_\theta$. Replacing $\mu_\theta(x,\psi)$ by the i.i.d.\ group mean $\bar r$ introduces an $O(1/G)$ ratio-estimation bias that vanishes as $G\to\infty$, and taking expectation over $(x,\psi)$ completes the result.
\end{proof}

\paragraph{Comparison with binary MaxRL and concurrent extensions.} For a binary indicator $R=\1[y\text{ passes on }\psi]\in\{0,1\}$, $\mu_\theta(x,\psi)=p_\theta(x,\psi)$, and the objective reduces to $\E_{x,\psi}\log p_\theta(x,\psi)$. Our formula gives $\nabla\log p_\theta(x,\psi)=\E_\theta[R\cdot S]/p_\theta(x,\psi)=\E_\theta[S\mid R=1]$, matching the MaxRL gradient of Tajwar et al.~\cite{tajwar2026maxrl} for fixed $(x,\psi)$. A concurrent extension, ArgMaxRL~\cite{doubleai2026argmaxrl}, instead decomposes graded rewards into threshold events to target best-of-$k$ performance; both reduce to MaxRL for binary rewards.

\subsection{Latent-success lifting}

\begin{proposition}
\label{prop:lift}
For fixed $(x,\psi)$, let $R_{\max}>0$ be an upper bound on $R(x,y;\psi)$ and define a latent Bernoulli variable $Z\mid(y,x,\psi)\sim\mathrm{Bernoulli}(R(x,y;\psi)/R_{\max})$. Then
\begin{equation}
\log\mu_\theta(x,\psi) = \log p_Z + \log R_{\max},\quad p_Z=\E_\theta[R/R_{\max}]=\E_{\theta,Z}[Z].
\end{equation}
Applying binary MaxRL to the $G$ lifted samples $(y_i,Z_i)$ gives the per-$(x,\psi)$ objective
\begin{equation}
J_G(\theta;x,\psi)=-\sum_{k=1}^{G}\frac{(1-\mu_\theta(x,\psi)/R_{\max})^k}{k},
\end{equation}
which converges to $\log\mu_\theta(x,\psi)$ (up to an additive constant $\log R_{\max}$) as $G\to\infty$ and satisfies $\nabla_\theta\E_{x,\psi}[J_G]=\nabla_\theta J_{\mathrm{MaxRL}}+O(1/G)$.
\end{proposition}
\begin{proof}
Since $p_Z=\mu_\theta(x,\psi)/R_{\max}$, $\log\mu_\theta(x,\psi)=\log R_{\max}+\log p_Z$. Truncating $-\sum_{k\ge1}(1-p_Z)^k/k=\log p_Z$ at $k=G$ yields the finite-$G$ MaxRL objective of Tajwar et al.~\cite{tajwar2026maxrl}, whose gradient is $\sum_{k=1}^G(1-p_Z)^{k-1}\nabla p_Z=\big(1-(1-p_Z)^G\big)\nabla\log p_Z\to\nabla\log p_Z=\nabla\log\mu_\theta(x,\psi)$ as $G\to\infty$. Substituting the conditional expectations under $Z\mid R$ into the numerator and denominator of the lifted advantage $A_i=(Z_i-\bar Z)/\bar Z$ (or taking $G\to\infty$) recovers the continuous-reward estimator.
\end{proof}

\subsection{Sensitivity to reward offset}

\begin{proposition}
\label{prop:shift}
For a shifted reward $R'=R+a$ with constant $a>0$, $J_{\mathrm{MaxRL}}(\theta,R')=\E_{x,\psi}\log(\mu_\theta(x,\psi)+a)$, and for each $(x,\psi)$
\begin{equation}
\nabla\log(\mu_\theta+a)=\frac{\nabla\mu_\theta}{\mu_\theta+a}=\frac{\mu_\theta}{\mu_\theta+a}\,\nabla\log\mu_\theta.
\end{equation}
As $a\to\infty$ the gradient approaches $(1/a)\nabla\E_{x,\psi}[\mu_\theta]$ (standard expected-reward RL). For any $a>0$, the per-$(x,\psi)$ weight becomes $1/(\mu_\theta+a)$ rather than $1/\mu_\theta$, strictly attenuating the MaxRL emphasis on hard inputs.
\end{proposition}
\begin{proof}
$\nabla\log(\mu_\theta+a)=\nabla\mu_\theta/(\mu_\theta+a)$. By the score-function identity, $\nabla\mu_\theta=\E_\theta[R\,S]$, so
\[
\nabla\log(\mu_\theta+a)=\frac{\E_\theta[R\,S]}{\mu_\theta+a}=\frac{\mu_\theta}{\mu_\theta+a}\cdot\frac{\E_\theta[R\,S]}{\mu_\theta}.
\]
The coefficient $\mu_\theta/(\mu_\theta+a)\in(0,1)$ shrinks the MaxRL gradient toward the standard RL gradient, which corresponds to $a\to\infty$ (where $\nabla\log(\mu_\theta+a)\approx\nabla\mu_\theta/a$).
\end{proof}

\paragraph{Implication for reward design.} \Cref{prop:shift} shows why the reward \cref{eq:reward} assigns failures a small positive value rather than zero or a large negative value. If failures score zero, a group of all-failures gives $A_i=0$ and the policy receives no gradient. If failures score a large constant, MaxRL degrades toward standard RL. The failure scale $s=0.2$ provides a fidelity-proportional gradient on hard inputs without washing out the MaxRL weighting. GRPO is insensitive to additive shifts (the group mean cancels), so this constraint applies only to MaxRL and Hybrid.

\section{Hybrid Advantage Estimator: Analysis}
\label{app:hybrid}

Fix an input $x$ and shared verifier state $\ket{\psi}$ with pass probability $p=p_\theta(x,\psi)>0$, and let $\mathcal E_G$ denote the event that all $G$ sampled candidates pass on $\psi$, which occurs with probability $p^G$. Conditioned on $\mathcal E_G$, the $G$ candidates are drawn i.i.d.\ from the conditional valid policy $\pi^+_\theta(y\mid x,\psi)=\pi_\theta(y\mid x)\1[y\text{ passes}]/p$. Because $r_i=b+g_i$ on $\mathcal E_G$, the bonus $b$ cancels in $A_i^{\mathrm{GRPO}}=(g_i-\bar g)/(\sigma_g+\varepsilon)$. Moreover, $\nabla_\theta\log\pi^+_\theta(y\mid x,\psi)=\nabla_\theta\log\pi_\theta(y\mid x)-\nabla_\theta\log p$, and the constant $\nabla_\theta\log p$ cancels against the zero-sum weights $\sum_{i=1}^G(g_i-\bar g)=0$. As $G$ grows, $(\bar g,\sigma_g)$ concentrates around the conditional mean $\mu^+_g(x,\psi)=\E_\theta[g(y)\mid y\text{ passes}]$ and standard deviation $\sigma^+_g(x,\psi)$, so the single-group GRPO gradient $\hat g^{\mathrm{GRPO}}=\frac{1}{G}\sum_{i=1}^G A_i^{\mathrm{GRPO}}\nabla_\theta\log\pi_\theta(y_i\mid x)$ converges in conditional expectation to the normalized gate-reduction gradient $\nabla_\theta\mu^+_g(x,\psi)/(\sigma^+_g(x,\psi)+\varepsilon)$ (with the threshold $\sigma_g\le\sigma_{\min}$ zeroing out updates once gate scores saturate).

Across both branches of \cref{eq:adv-hybrid}, total expectation decomposes the single-group Hybrid gradient as
\begin{multline}
\E_\theta\big[\hat g^{\mathrm{Hybrid}}\,\big|\,x,\psi\big]
= \big(1-p^G\big)\,\E_\theta\big[\hat g^{\mathrm{MaxRL}}\,\big|\,\mathcal E_G^c,x,\psi\big]\\
+ p^G\,\E_\theta\big[\hat g^{\mathrm{GRPO}}\,\big|\,\mathcal E_G,x,\psi\big].
\label{eq:hybrid-decomp}
\end{multline}
When an input is not yet reliably solved ($p^G\ll 1$; e.g., $p\le 0.8$ gives $p^{32}<10^{-3}$), $\Pr(\mathcal E_G^c\mid x,\psi)=1-p^G\approx 1$ and the first term matches the unconditional continuous MaxRL gradient of Appendix~\ref{app:maxrl} up to $O(p^G)$, prioritizing equivalence. As $p\to 1$ ($p^G\to 1$), the second term dominates and updates transition smoothly to GRPO gate-count optimization without additional schedule hyperparameters.

\section{Theoretical Basis of Random-State Verification}
\label{app:fidelity}

This appendix gives the theoretical basis of the verifier of \S\ref{sec:verify}. Throughout, $n$ is the number of qubits, $d=2^n$, and $W=U(\Cr)^\dagger U(\Cgen)$. For a state $\ket\psi$, writing $F(\psi)\equiv F(\Cgen;\psi)$, we define
\begin{equation}
F(\psi)=|\bra\psi W\ket\psi|^2,\qquad X(\psi)=1-F(\psi)\in[0,1],
\end{equation}
and we compare the two circuits through the process fidelity
\begin{equation}
F_{\mathrm p}=\frac{|\Tr W|^2}{d^2},\qquad r_{\mathrm p}=1-F_{\mathrm p}.
\end{equation}
Since $|\Tr W|\le d$, with equality if and only if all eigenvalues of $W$ coincide, the two circuits are equivalent up to a global phase exactly when $F_{\mathrm p}=1$.

\subsection{Single-state fidelity estimates average state fidelity}

The verifier draws $\ket\psi$ from the Haar measure. The second moment of this measure is $\E_\psi\big[(\ket\psi\!\bra\psi)^{\otimes2}\big]=(I+S)/\big(d(d+1)\big)$, where $S$ swaps the two tensor factors~\cite{mele2024introduction}. With $\Tr[(A\otimes B)S]=\Tr(AB)$, the average fidelity is
\begin{equation}
\bar F=\E_\psi[F(\psi)]=\frac{\Tr(WW^\dagger)+|\Tr W|^2}{d(d+1)}=\frac{d+|\Tr W|^2}{d(d+1)}.
\end{equation}
Substituting $|\Tr W|^2=d^2F_{\mathrm p}$ gives
\begin{equation}
\bar F=\frac{d\,F_{\mathrm p}+1}{d+1},\qquad r_{\mathrm{avg}}=1-\bar F=\frac{d}{d+1}\,r_{\mathrm p}.
\label{eq:fid-relation}
\end{equation}
For any candidate fixed prior to drawing $\ket\psi$, the fidelity on one Haar-random state is therefore an unbiased estimate of the average state fidelity $\bar F$, and its infidelity $X(\psi)$ is an unbiased estimate of $r_{\mathrm{avg}}$, which equals the process infidelity up to the factor $d/(d+1)\in[32/33,1)$ for $n\ge5$. Errors in the two quantities are tied directly: a candidate is close to the reference on average over input states if and only if its process fidelity is close to one.

\subsection{Statistical guarantee of multi-state re-verification}

Consider a multi-state test that draws $k$ independent Haar-random test states and accepts a fixed candidate only if it passes on all $k$ states. We bound the probability of accepting a fixed candidate whose average infidelity is non-negligible. Diagonalize $W$ with eigenvalues $\lambda_1,\dots,\lambda_d$, $|\lambda_i|=1$. By unitary invariance of the Haar measure, the weights $p_i=|\braket{i}{\psi}|^2$ in the eigenbasis are uniformly distributed on the simplex, $(p_1,\dots,p_d)\sim\mathrm{Dirichlet}(1,\dots,1)$. With $a_{ij}=|\lambda_i-\lambda_j|^2\ge0$,
\begin{equation}
X=1-\Big|\sum_i p_i\lambda_i\Big|^2=\sum_{i<j}a_{ij}\,p_ip_j .
\end{equation}
The Dirichlet moments $\E[p_ip_j]=1/\big(d(d+1)\big)$ for $i\ne j$ give $r_{\mathrm{avg}}=\sum_{i<j}a_{ij}/\big(d(d+1)\big)$. For $i<j$ and $u<v$, each fourth moment $\E[p_ip_jp_up_v]$ appearing in $X^2$ is at most $4/\big(d(d+1)(d+2)(d+3)\big)$, so
\begin{equation}
\E[X^2]\le\kappa_d\,r_{\mathrm{avg}}^2,\qquad \kappa_d=\frac{4d(d+1)}{(d+2)(d+3)}<4.
\end{equation}
Since $X\ge0$, the Paley--Zygmund inequality gives, for $0\le\tau<r_{\mathrm{avg}}$,
\begin{equation}
\Pr_\psi[X>\tau]\ \ge\ \frac{(1-\tau/r_{\mathrm{avg}})^2\,r_{\mathrm{avg}}^2}{\E[X^2]}\ \ge\ \frac{(1-\tau/r_{\mathrm{avg}})^2}{\kappa_d}.
\end{equation}
A single random-state test accepts a candidate when $X\le\tau_F$. If $r_{\mathrm{avg}}\ge2\tau_F$, each test therefore rejects with probability at least $(1/2)^2/4=1/16$. Across $k$ independent random states, all of which must pass,
\begin{equation}
\Pr[\text{accept}]\le(15/16)^k\qquad\text{whenever } r_{\mathrm{avg}}\ge2\tau_F.
\label{eq:false-accept}
\end{equation}
This worst-case bound holds for any fixed unitary satisfying $r_{\mathrm{avg}}\ge 2\tau_F$ at every $n$, so $k\ge\ln\delta/\ln(15/16)$ states suffice to drive the bound below $\delta$ (e.g., $k=108$ for $\delta=10^{-3}$ and $k=215$ for $\delta=10^{-6}$). By \cref{eq:fid-relation}, any fixed candidate whose process infidelity is at least $2\tau_F(d+1)/d$ is rejected by $k$ independent random states with probability at least $1-(15/16)^k$.

\subsection{Infidelity of discrete Clifford+\texorpdfstring{$T$}{T} errors}
\label{app:fidelity-gap}

To contextualize the empirical process-infidelity distribution observed across the 32{,}000 candidates in \S\ref{sec:res-verify}, we distinguish single-gate edits from multi-gate errors over $\mathcal G$:

\begin{enumerate}
  \item \emph{Exact lower bound for single-gate edits.} Suppose $\Cgen$ differs from $\Cr$ by inserting, deleting, or substituting a single gate at an arbitrary position in the circuit, $\Cr=(C_{\mathrm{before}},\allowbreak G_1,\allowbreak C_{\mathrm{after}})$ and $\Cgen=(C_{\mathrm{before}},\allowbreak G_2,\allowbreak C_{\mathrm{after}})$ for inequivalent $G_1,G_2\in\mathcal G\cup\{I\}$. Writing $A=U(C_{\mathrm{after}})$ and $B=U(C_{\mathrm{before}})$, we have $U(\Cr)=A\,U(G_1)\,B$, $U(\Cgen)=A\,U(G_2)\,B$, and $W=B^\dagger U(G_1)^\dagger A^\dagger A\,U(G_2) B = B^\dagger U(G_1)^\dagger U(G_2) B$. Both the shared suffix $C_{\mathrm{after}}$ and, by cyclic invariance of the trace, the shared prefix $C_{\mathrm{before}}$ cancel in $F_{\mathrm p}=|\Tr(U(G_1)^\dagger\allowbreak U(G_2))|^2\allowbreak /d^2$, regardless of circuit depth, entanglement, or whether $C_{\mathrm{after}}$ commutes with the edited gate. Because any pair $(G_1,G_2)$ acts on at most four qubits and identity factors on untouched qubits cancel, finite enumeration gives the exact minimum non-zero process infidelity
  \begin{equation}
  r_{\mathrm p,\min}^{(1)}=1-\frac{|1+e^{\pm i\pi/4}|^2}{4}=\sin^2\!\left(\frac{\pi}{8}\right)\approx 0.1464,
  \end{equation}
  attained by an uncancelled $\pm\pi/4$ ($T$ or $T^\dagger$) phase difference on one qubit (followed by $0.2714$, $0.5000$ for a single $S$ edit, $0.7500$ for an inserted/deleted CNOT or CZ, and $1.0$ for $H$, $X$, or $Z$).

  \item \emph{Multi-gate errors and empirical distribution.} Because Clifford+$T$ generates a dense subgroup of $\mathrm{PU}(2^n)$, $r_{\mathrm p,\min}^{(1)}$ is not a universal lower bound when multiple gates differ: for instance, reversing two non-commuting $\pi/4$ rotations on one qubit ($W=T^\dagger(HT^\dagger H)\allowbreak T(HTH)$) yields $r_{\mathrm p}=1-\big(1-2\sin^4(\pi/8)\big)^2\approx 0.0839$; an $m$-qubit controlled-$Z$ discrepancy $C^{m-1}Z$ yields $r_{\mathrm p}=1-(1-2^{1-m})^2\approx 2^{2-m}$ ($0.1211$ at $m=5$, falling below $2\tau_F=2\times 10^{-4}$ only for $m\ge 15$); and long Solovay--Kitaev sequences can approach zero arbitrarily closely. Conversely, when an early erroneous token causes an autoregressive policy to generate a divergent continuation ($C'_{\mathrm{after}}\ne C_{\mathrm{after}}$), multiple gates differ without symmetric cancellation, explaining why no inequivalent candidate in \S\ref{sec:res-verify} falls below $0.1464$ ($89.4\%$ have $r_{\mathrm p}\ge 0.9$).
\end{enumerate}

\subsection{Use in training and evaluation}

Training and fast decoding use a single random input state per group ($k=1$) with $\tau_F=10^{-4}$; all $G$ candidates in a group share this state, evaluating candidates on a common input while reusing a single evolution of $\Cr$, and each candidate fixed prior to drawing $\ket\psi$ retains $\bar F$ as its marginal expected fidelity. For final evaluation (\S\ref{sec:res-verify}), all $80{,}674$ selected best candidates across all reported model--bucket evaluations are deterministically certified against $\Cr$ via exact unitary comparison for $n\le 10$ ($|\Tr(U(\Cr)^\dagger U(\Cgen))|/d\ge 1-10^{-8}$; $24{,}910$ candidates) and symbolic ZX-calculus equivalence verification (\texttt{pyzx.\allowbreak Circuit.\allowbreak verify\_equality}) across all $5\le n\le 20$.

\section{Additional Experimental Results}
\label{app:results}

\begin{figure*}[t]
\centering
\includegraphics[width=\textwidth]{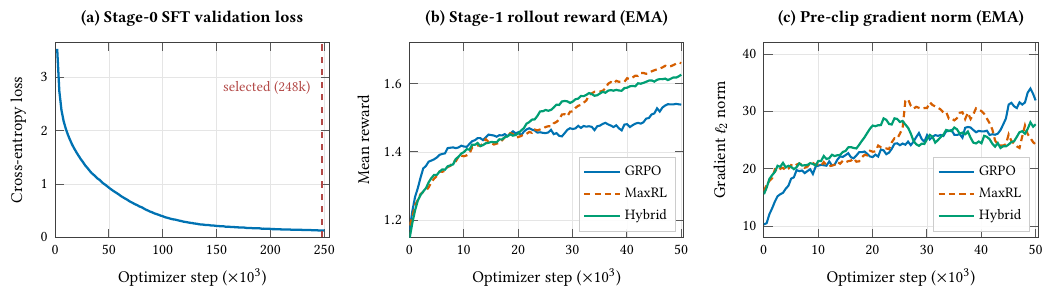}
\caption{Stage-0 and stage-1 training trajectories. \textbf{(a)} Stage-0 SFT validation cross-entropy loss, marking the step-248{,}000 checkpoint used to initialize RL. \textbf{(b)} Stage-1 mean rollout reward (exponential moving average, $\alpha=0.92$). \textbf{(c)} Stage-1 pre-clip gradient $\ell_2$ norm (EMA, $\alpha=0.95$).}
\label{fig:app-train}
\end{figure*}

\begin{figure*}[t]
\centering
\includegraphics[width=\textwidth]{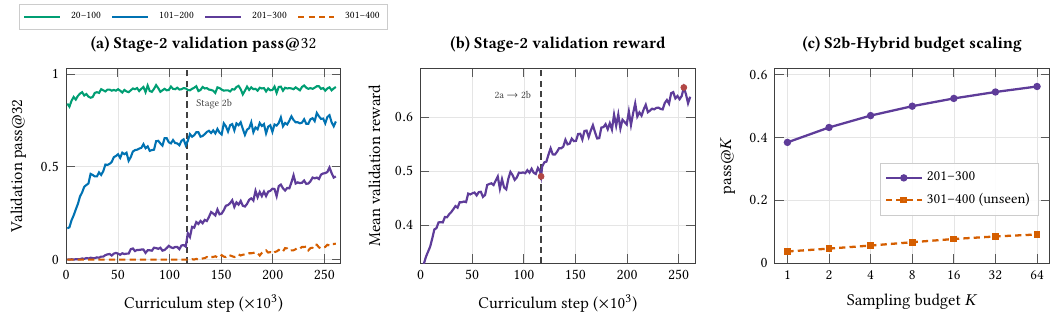}
\caption{Stage-2 curriculum dynamics and test-time scaling. \textbf{(a)} Per-bucket validation pass@$32$ across Stage 2a (steps 0--116{,}736) and Stage 2b (steps 116{,}736--262{,}144). \textbf{(b)} Mean validation reward, with red markers indicating S2a-Hybrid (116.7k) and S2b-Hybrid (255.0k). \textbf{(c)} Test pass@$K$ ($K\in\{1,\dots,64\}$) of S2b-Hybrid (step 254{,}976) on the 992-circuit stratified subsets of 201--300 ($38.5\%\to 56.3\%$) and unseen 301--400 ($3.8\%\to 9.2\%$) gates.}
\label{fig:app-s2}
\end{figure*}

\begin{figure*}[t]
\centering
\includegraphics[width=\textwidth]{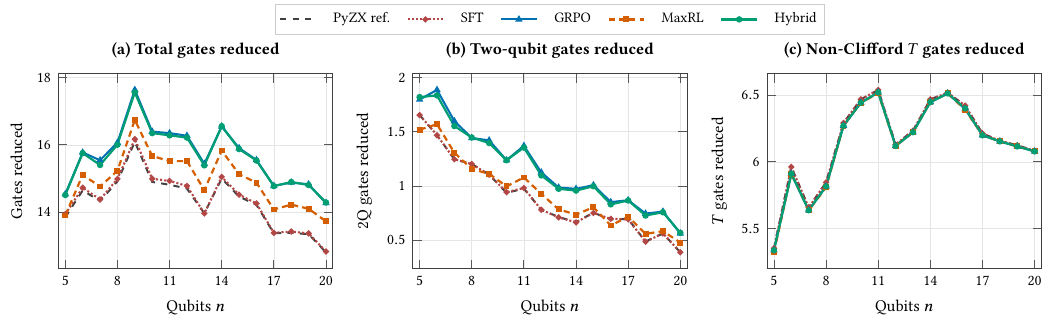}
\caption{Stage-1 per-qubit gate reductions relative to $\Csrc$ on the common success set ($n_{\mathrm{com}}=5{,}443$, 20--100-gate test split, $K=32$, $\mathcal T=0.2$). \textbf{(a)} Total gates reduced. \textbf{(b)} Two-qubit gates reduced. \textbf{(c)} Non-Clifford $T$ gates reduced.}
\label{fig:app-est-gates}
\end{figure*}

\Cref{fig:app-train,fig:app-s2,fig:app-est-gates} present supplementary training trajectories, test-time scaling curves on longer buckets, and per-qubit gate-reduction breakdowns complementing \S\ref{sec:results}.

\Cref{fig:app-train} shows the stage-0 supervised validation loss alongside stage-1 rollout rewards and pre-clip gradient norms; across GRPO, MaxRL, and Hybrid, pre-clip gradient norms remain within a comparable bounded range ($15$--$35$) prior to unit norm clipping (which effectively normalizes step sizes across estimators). \Cref{fig:app-s2}(a,b) plots validation accuracy and reward across the full 262{,}144-step stage-2 curriculum, showing the immediate inflection on 201--300 and unseen 301--400 gates at the Stage-2a$\to$2b transition (step 116{,}736), while \cref{fig:app-s2}(c) reports $K\in\{1,\dots,64\}$ budget scaling of S2b-Hybrid on the 992-circuit stratified subsets of 201--300 and 301--400 gates. Finally, \cref{fig:app-est-gates} gives the per-qubit breakdown ($5\le n\le 20$) of total, two-qubit, and $T$-gate reductions across all stage-1 policies on the common success set of \cref{tab:estimator-results}.

\end{document}